\documentclass[aip,
amsmath,amssymb,
reprint, %
]{revtex4-1}

\usepackage{graphicx}
\usepackage{dcolumn}
\usepackage{bm}

\begin{document}

\preprint{AIP/123-QED}

\title{Phase ordering of zig-zag and bow-shaped hard needles in two dimensions}


\author{Raffaele Tavarone}
\affiliation{Institut f\"ur Theoretische Physik, Technische Universit\"at Berlin,
Hardenbergstrasse 36, D-10623 Berlin, Germany}

\author{Patrick Charbonneau}
\email{patrick.charbonneau@duke.edu}
\affiliation{Departments of Chemistry and Physics, Duke University, Durham, North Carolina 27708, USA}

\author{Holger Stark}
\email{holger.stark@tu-berlin.de}
\affiliation{Institut f\"ur Theoretische Physik, Technische Universit\"at Berlin,
Hardenbergstrasse 36, D-10623 Berlin, Germany}

%


\begin{abstract}
We perform extensive Monte Carlo simulations of a two-dimensional bent hard-needle model in both its chiral zig-zag and its achiral bow-shape configurations
and present their phase diagrams.
We find evidence for a variety of stable phases: isotropic, quasi-nematic, smectic-C, anti-ferromorphic smectic-A, and modulated-nematic. This last phase 
consists of layers formed by supramolecular arches. They create a periodic modulation of the
molecular polarity 
whose period 
is sensitively controlled by molecular geometry. 
We identify transition densities
using correlation functions
together with
appropriately defined
order parameters
and 
compare them
with predictions from Onsager theory. The contribution of 
the 
molecular
excluded area 
to deviations from Onsager theory and simple liquid crystal phase morphology is discussed. 
We demonstrate the
isotropic--quasi-nematic transition 
to be
consistent with a Kosterlitz-Thouless disclination unbinding scenario.
\end{abstract}

\maketitle

\section{Introduction}

In recent years, interest in two-dimensional and quasi-two-dimensional self-assembled structures in thin films has grown tremendously\cite{ulman2013introduction,schreiber2004self,boker2007self}. The possibility to fine-tune molecular ordering indeed makes thin films suitable for myriad technological applications\cite{barth2005engineering}, ranging from electronics\cite{aswal2006self} and optics\cite{hicks1999nonlinear} to biology\cite{mendes2008stimuli}. The formation of liquid-crystal phases on two-dimensional surfaces is also key for various nanotechnological applications \cite{hore2010nanorod,mclean2006controlled,slyusarenko2014two}.

Control over self-assembly has already enabled the formation of two-dimensional aggregates with quasicrystal\ \cite{fournee2014self,mikhael2010proliferation,schmiedeberg2008colloidal,schmiedeberg2010archimedean}, hexagonal\ \cite{he2005self}, crystal\ \cite{winfree1998design}, and liquid crystal\ \cite{bai2010recent,de2003structure,muvsevivc2006two} orders. Improving control on the spontaneous formation of ordered structures, however, requires a deep understanding 
of how molecular geometry
influences supramolecular ordering. A common starting point for 
studying this relation are hard-core models, which recapitulate the static and dynamical properties of a wide range of phenomena, from the hard-sphere-like freezing of atomic liquids \cite{mulero2008theory} to quasicrystal formation~\cite{haji2009disordered}.  These models also reproduce the rich two-dimensional liquid crystal ordering of objects with high aspect ratios such as rods\cite{Bates2000,ghosh2007orientational,kahlitz2012phase}, rectangles\cite{donev2006tetratic,martinez2005effect}, spherocylinders\cite{lagomarsino2003isotropic,kahlitz2012clustering}, and ellipsoids\cite{xu2013hard,moradi2010monte}. The good agreement between hard-core models and some experiments suggests that entropy alone can suffice to drive the formation of ordered structures\cite{care2005computer}.
 
High-aspect ratio molecules with a bent core\cite{Lubensky2002,Takezoe2006} or a banana shape\cite{Pelzl1999,Ros2005} 
assemble in an even richer set of morphologies, including biaxial nematic\cite{Berardi2008,Teixeira1998}, fan-shaped texture\cite{Pelzl2002}, and nematic with splay-bend deformation or conical twist-bend helix\cite{Dozov2001a}.
From the technological viewpoint these structures are quite interesting. Biaxial nematic phases, for instance, 
exhibit ferro- or antiferro-electric properties\cite{Takezoe2006}. A plethora of models have thus been devised 
to understand the three-dimensional bulk behavior of these systems
\cite{Camp1999a,johnston2002computer,dewar2004computer,shamid2013statistical,grzybowski2011biaxial}. Yet relatively little theoretical attention has been paid to related models in two dimensions\cite{Martinez-Gonzalez2012,Bisi2008,Martinez-Gonzalez2012a}.

Here, we consider the two-dimensional phase behavior of a bent-needle model in both its  chiral zig-zag and achiral bow-shaped configurations (Fig.\ \ref{fig: model}). These two versions display significantly different mesophases,
for which we map out complete phase diagrams.
Zig-zag molecules are known to assemble in either a nematic or a smectic-C phase depending on the packing density $\rho$\cite{Peon2006}, and bow-shaped molecules have been found to display tetradic and nematic order\cite{Martinez-Gonzalez2012}. 
Yet these studies did not clarify the role of topological defects and of thermal fluctuations on the long-range stability of the mesophases, which are of fundamental physical interest\cite{PhysRev.176.250,strandburg1988two,bernard2011two}. 
The relatively coarse sampling of configuration space previously used further left open the possibility that qualitative features of the molecular ordering may have been missed. We indeed find that in addition to forming quasi-nematic and smectic phases, bow-shaped molecules present a stable modulated-nematic phase, as was previously predicted\cite{Dozov2001a} and observed \cite{borshch2013nematic,memmer2002liquid} in three-dimensional systems.  

Our advances are not only made possible by the use of specialized Monte Carlo simulations on large systems, but also by the definition of appropriate order parameters and correlation functions. Our analysis thence extends 
investigations of
two-dimensional systems of hard needles\cite{kahlitz2012phase,Frenkel1985,Vink2009}, 
hard spherocylinders\cite{Bates2000,kahlitz2012clustering}, and spherocylinders with a polar head\cite{Armas-Perez2011}.     
These improved numerical results 
on two-dimensional liquid crystal formation are also compared to predictions from
Onsager theory\cite{Varga2009b}.

The paper is organized as follows. In Sec.\ \ref{sec: Model} we describe the model and provide details on the 
Monte Carlo simulation procedure. Section\ \ref{sec.quasi} reviews the theory of quasi-long-range orientational order 
in two dimensions. In Sec. \ref{sec: result} we present the results of our analysis of the simulation data. 
In Sec. \ref{sec: Onsager Theory} we summarize the Onsager Theory for two-dimensional hard-core objects and apply 
it to the bent-hard needle model.
We conclude in Sec.\ \ref{sec: conclusion}.


\section{\label{sec: Model}Model and Simulation Methods}

\subsection{Bent hard needles}
\begin{figure}
\begin{center}
\includegraphics{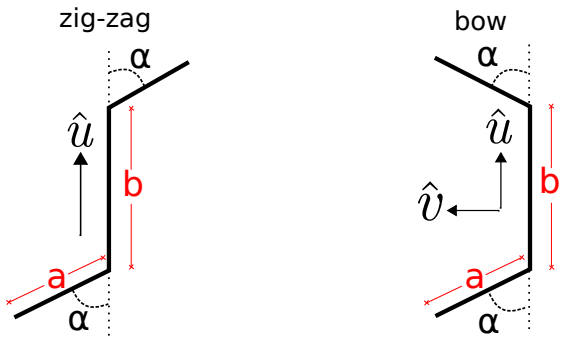}
\caption{Bent-needle model molecule in chiral zig-zag (left) and achiral bow-shaped (right) configurations.
}
\label{fig: model}
\end{center}
\end{figure}

Our
bent hard-needle model
consists of a central line segment of length $b$ to which
two terminal line segments of equal length $a$ are attached at a fixed angle $\alpha$  (Fig.\ \ref{fig: model}). The molecules can 
adopt a chiral zig-zag or an achiral bow-shape configuration, depending on whether $\alpha$ is defined on the same or on opposite sides of the central segment. Note, however, that we only consider enantiomerically pure systems in order to avoid chiral segregation\cite{Perusqua2005}.
Pairs of molecules interact via a hard-core exclusion potential but are infinitely thin, 
i.e., they can be infinitely close to one another but cannot overlap. 
Since the hard-core interaction potential is athermal, we set to unity the product of Boltzmann's constant and temperature, $k_{\mathrm{B}}T=1/\beta=1$, without loss of generality.
The total length of the molecule,
$L = 2a + b=1$, is used as the unit of length, also without loss of generality. Each model is thus completely determined by 
two parameters: $a$ and $\alpha$.  In the following we let $\alpha$ vary from 0 to $\pi/2$ (at $\alpha=0$ both models are equivalent), but we fix $a=0.25$ for zig-zag molecules and $a=0.35$ for bow-shaped molecules. This choice
maximizes the excluded area (to be defined below), and thus pushes down the isotropic--mesophase transition densities,
making them computationally less costly to study.

\begin{figure}
\begin{center}
\includegraphics{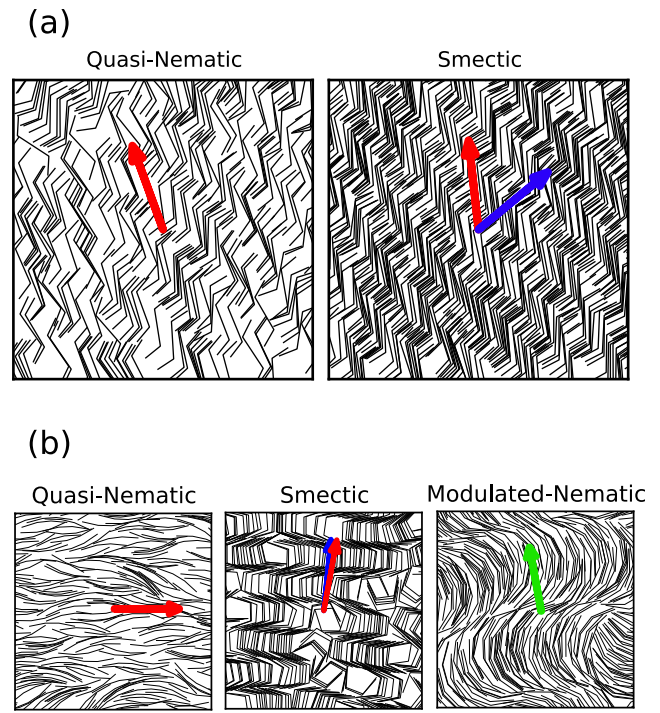}
\caption{(colors online) Snapshots  of the main mesophases identified in Monte Carlo simulations: (a) zig-zag molecules with $a=0.25$, $\alpha=\pi/3$ in both the quasi-nematic ($\rho=14$) and the smectic ($\rho=30$) phases, and (b) bow-shaped molecules with $a=0.35$, $\alpha=\pi/16$, $\rho=20$  in the quasi-nematic, with $a=0.35$, $\alpha=\pi/3$, $\rho=30$ in the smectic, and with $a=0.35$, $\alpha=\pi/6$, $\rho=26$  in the modulated-nematic phases. The nematic director, the smectic surface normal and the modulated-nematic layers normal are indicated as red, blue, and green arrows, respectively. Note that in order to clearly illustrate the mesophase morphology, only a portion of the simulation box is displayed.
}
\label{fig: configuration}
\end{center}
\end{figure}

Figure\ \ref{fig: configuration} depicts the quasi-nematic and smectic phases for zig-zag and bow-shaped molecules as well as the modulated-nematic phase of bow-shaped molecules. This last phase, which consists of layers made up of arches formed by several molecules\cite{Dozov2001a}, is discussed in more details in
Sec.\ \ref{subsec: supra layering}.

\subsection{Order Parameters} 
We define the two-dimensional nematic order parameter $S$ by first introducing the
tensor order parameter $\boldsymbol{Q}$:
\begin{equation}
Q_{\alpha \beta } = \langle 
N^{-1}\sum_{i=1}^{N} (2u_{\alpha}^{i}u_{\beta}^{i}-\delta_{\alpha \beta})
\rangle \, ,
\end{equation}
where $u_{\alpha}^{i}$ is the $\alpha$-th Cartesian coordinate of the unit vector 
pointing along the central segment of the $i$-th of $N$ molecules and $\langle \ldots \rangle$ denotes the ensemble average. The positive eigenvalue and corresponding eigenvector of $\mathbf{Q}$ give $S$ and the nematic director $\bm{n}$, respectively. The nematic order parameter $S$ is canonically used to quantify the degree of molecular alignment along $\bm{n}$ -- perfect alignment has $S=1$. In two dimensions, however, only quasi-long-range orientational, i.e., quasi-nematic, order can exist (see Sec.~\ref{sec.quasi}), and hence 
$S=0$ 
in the thermodynamic limit, $N \to \infty$ \cite{Frenkel1985}. 
In order to characterize orientational order in the quasi-nematic phase, we resort to the orientational correlation function 
\begin{equation}
g_2(r)=\langle \cos[2(\theta(0)-\theta(r))]\rangle \, ,
\label{eq: orientational correlation}
\end{equation}
where $\theta$ is the angle between the
central molecular
segment
along $\hat{\bm{u}}$
and a fixed axis, and $r$ is the distance between the centers of two molecules. 
The function $g_2(r)$ thus monitors the spatial decay of orientational correlations.
Note that because we work under periodic boundary conditions, correlations are radially truncated at half the edge length of the simulation box.

\begin{figure}
\begin{center}
\includegraphics{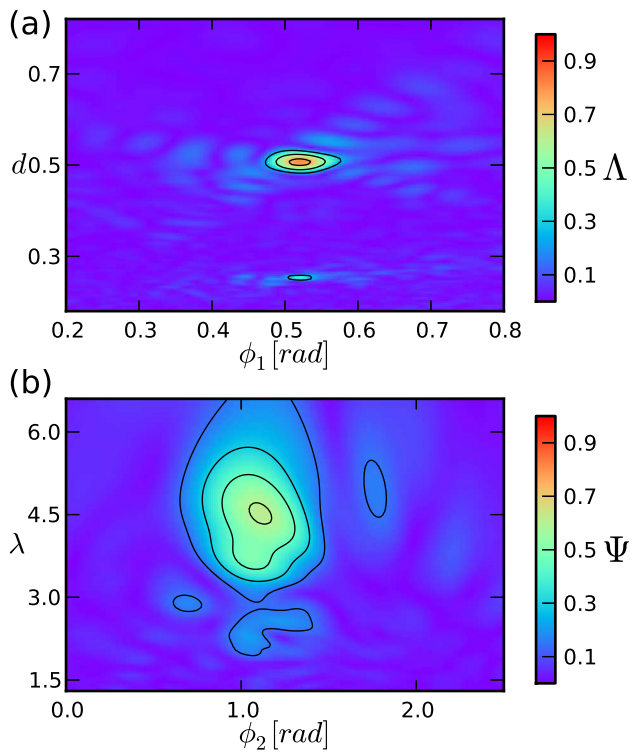}
\caption{(colors online) 
Illustration of the maximization procedure for determining the smectic (a) and modulated-nematic (b) order parameter.
(a) Fourier transform of the density, $\Lambda(\hat{\bm{k}},d)$, plotted versus orientation angle $\phi_1$  and 
wavelength $d$ of the wave vector $\bm{k} = (2\pi/d) \hat{\bm{k}}$ with $\hat{\bm{k}}=[\cos \phi_1,\sin \phi_1 ]$.
The data refer to
zig-zag molecules with $a=0.25$, $\alpha=\pi/2$, and $\rho=30$. The maximization gives
 $\Lambda_{\mathrm{sm}}=0.83$.
(b) Fourier transform of the polar order parameter, $\Psi(\hat{\bm{h}},\lambda)$, plotted versus orientation angle $\phi_2$ 
and wavelength $\lambda$ of the wave vector $\bm{h}=(2\pi/\lambda)\hat{\bm{h}}$ with $\hat{\bm{h}}=[\cos \phi_2,\sin \phi_2 ]$.
The data refer to bow-shaped molecules with $a=0.35$, $\alpha = \pi/13$, $\rho = 24$. The maximization 
gives $\Psi_{\mathrm{m}}=0.62$. 
In both cases the procedure gives rise to a clear maximum, which is used for further analysis. Note that only a portion of the domain explored during the optimization procedure is displayed here.
}
\label{fig: Fourier Maximization}
\end{center}
\end{figure}

A smectic liquid crystal can be thought of as a stack of parallel molecular layers of thickness $d$. It is therefore possible to identify a density wave along the normal to the layers. The smectic order parameter $\Lambda_{\mathrm{sm}}$ is then the amplitude of this density wave.
To determine
$\Lambda_{\mathrm{sm}}$,
we calculate the Fourier transform of
the normalized density
\cite{Polson1997},
\begin{equation}
\Lambda(\hat{\bm{k}},d) =
\Bigg \langle \frac{1}{N}   
\Bigg|
\sum_{j=1}^{N} e^{i\bm{k}\cdot \bm{r}_{j}} 
\Bigg|
\Bigg\rangle \, ,
\label{eq: smectic order parameter}
\end{equation}
where $\bm{r}_{j}$ is the position of the $j$-th particle, and $\bm{k}= (2\pi/d)\hat{\bm{k}}$ is the wave vector with 
wavelength
$d$ and 
unit vector $\hat{\bm{k}}$.
To determine the actual distance $d$ between smectic layers and the layer orientation $\hat{\bm{k}}$,
we calculate
$\Lambda(\hat{k},d)$ for different values of $d$ and directions $\hat{\bm{k}}$, and take its maximum $\Lambda_{\mathrm{sm}}$
as the smectic order parameter\cite{Peon2006}.
Figure\ \ref{fig: Fourier Maximization}(a) illustrates this optimization procedure.

Bow-shaped molecules are able to display polar order. To quantify it, we introduce
the global polar order parameter 
\begin{equation}
P_{\mathrm{g}}= \left\vert \frac{1}{N}\sum_{i=1}^{N}\hat{\bm{v}}^{i}  \right\vert   \, ,
\end{equation}
where 
$\hat{\bm{v}}^{i}$ is a unit vector perpendicular to the molecular axis $\hat{\bm{u}}^{i}$
(see Fig.\ \ref{fig: model}). 
Since $P_{\mathrm{g}}$ cannot distinguish between isotropic and anti-polar 
order -- in both cases $P_{\mathrm{g}}=0$ --, other order parameters are needed to characterize 
an anti-polar state.
In Sec. \ref{subsec: supra layering} we describe the modulated-nematic phase \cite{Dozov2001a} that
bow-shaped molecules form
at intermediate $\alpha$.
In this mesophase molecules form
arches
in which the orientation of the molecules' central segments --
and thus also the polar vector $\hat{\bm{v}}$ (see Fig.\ \ref{fig: model}) --
display
a periodic modulation 
along one particular spatial direction. 
To quantifiy this spatial modulation, we
introduce its amplitude as an
order parameter 
$\Psi_{\mathrm{m}}$,
which we determine in full analogy to the definition of the smectic order parameter
$\Lambda_{\mathrm{sm}}$.
We define the spatially dependent polarity,
$P_{\mathrm{l}}(\bm{r})=\sum_{j=1}^{N} e^{i \theta_{j}} \delta(\bm{r}-\bm{r}_j)$, 
and calculate the amplitude of its Fourier transform
\begin{equation}
\Psi(\hat{\bm{h}},\lambda)=
\Bigg \langle \frac{1}{N}   
\Bigg|
\sum_{j=1}^{N} e^{i \theta_j} e^{i\bm{h}\cdot \bm{r}_{j}} 
\Bigg|
\Bigg\rangle \, ,
\label{eq: modulated order parameter}
\end{equation}
where $\bm{h}=(2\pi/\lambda)\hat{\bm{h}}$ is the wave vector with wavelength $\lambda$ and unit vector $\hat{\bm{h}}$. Note 
since
$\theta + \pi/2$ quantifies the orientation of $\hat{\bm{v}}$, $\Psi$ 
describes
periodic variations in $\hat{\bm{v}}$.
We evaluate $\Psi(\hat{\bm{h}},\lambda)$ for different values of $\lambda$ and directions $\hat{\bm{h}}$ and take the maximum 
$\Psi_{\mathrm{m}}$ as the 
order parameter
for the modulated-nematic phase.
The optimization procedure is illustrated in Fig.\ \ref{fig: Fourier Maximization}(b).        

In order to have 
additional information on
the structural properties of the modulated-nematic phase, 
we monitor polar correlations along the optimal direction $\hat{\bm{h}}_{\mathrm{m}}$
using the polar correlation function
\begin{equation}
g_{1}^{\vert \vert}(r)=
\Bigg \langle \frac{1}{N}   
\sum_{i=1}^{N}\sum_{j \neq i} \hat{\bm{v}}^{i} \cdot \hat{\bm{v}}^{j} \delta \Big( r^{\vert \vert}_{ij} - r \Big) 
\Bigg\rangle \, .
\label{eq: modulated-correlation-function}
\end{equation}
Here,
$r^{\vert \vert }_{ij} = \vert \bm{r}_{i} - \bm{r}_{j} \vert \cdot \hat{\bm{h}}_{\mathrm{m}}$ is the projection of the vector joining the centers of the $i$-th and $j$-th molecules
onto the
direction of $\hat{\bm{h}}_{\mathrm{m}}$.

\subsection{Monte Carlo simulations}

In order to study the phase behavior of the zig-zag and bow-shaped models, we 
perform Monte Carlo simulations 
mostly
in 
the constant-$NVT$ 
but also in the
constant-$NPT$ ensemble under periodic boundary conditions\cite{Frenkel1997}. 
Most simulations have $N = 2000$ in a square box of 
area $V$. 

Constant-$NVT$ simulations are performed at fixed reduced 
density $\rho=N/V$ and consist of up to $3.0\times 10^{7}$ sweeps, where
a Monte Carlo sweep 
comprises $N$ independent trial displacements.
The basic Monte Carlo move in the $NVT$ ensemble consists of either translating or rotating (randomly chosen with equal probability) a randomly chosen molecule. The maximum attempted displacement is preliminarily adjusted in such a way that an acceptance rate of approximately $50 \%$ is achieved, but during the production runs the maximum step sizes are fixed -- dynamical adjustments would violate detailed balance\cite{Miller2000}.
The hard-core nature of the pair interaction simplifies the Metropolis acceptance rule:
the displacement of a molecule is accepted if
it does not generate an overlap, and
is rejected otherwise.
In order to check that the sampled configurations correspond to equilibrium (and not metastable) states, different initial configurations -- including isotropic, perfect polar and anti-polar nematic, as well as perfect smectic states -- are used. Note that reaching equilibrium and properly sampling states sometimes require up to several million Monte Carlo sweeps, leading to 
computational times of up to four weeks on a single-thread of an Intel Xeon X5550 machine with a 2.66 GHz CPU.
The minimum simulation time was about one week.

Constant-$NPT$ simulations also include changes 
in $V$ in order to keep the system pressure $P$ constant\cite{Frenkel1997}. A Monte Carlo sweep then comprises an average of $N$ independent single-molecule trial displacements and one trial volume change. In 
a trial volume change,
$V$ is 
modified by an amount $\Delta \ln V$
and all the molecule positions are rescaled accordingly.
If no overlap occurs in the resulting configuration, the move is accepted with probability
\begin{equation}
p(V \rightarrow V+\Delta V)= \exp[-P\Delta V + (N+1)\ln (1+\Delta V/V)],
\end{equation}
and is rejected otherwise.

For the bow-shaped molecules with intermediate $\alpha$, constant-$NVT$ simulations either result in ferromorphic or in anti-ferromorphic
states, depending on the initial configuration. In order to identify the equilibrium ground state, we thus also performed a slow pressure annealing. This process was achieved through a sequence of constant-$NPT$ simulations
starting from the isotropic regime and increasing $P$ in steps of 10-20\%.
At each step $\rho$ was equilibrated
for at least $2.0\times 10^6$ sweeps.
The opposite procedure was used to check for hysteresis. 
This
study revealed that the anti-ferromorphic phase is the equilibrium state, as we discuss in Sec.\ \ref{subsec: supra layering}.

Since the interaction between bent needles is short-ranged, we use a combination of linked and Verlet lists 
to check for the overlap of molecules after a Monte Carlo trial. This speeds
up the simulations\cite{Frenkel1997,Sutmann2006}.
The Verlet list is built by wrapping a spherocylinder body around each molecule.
This non-trivial shape is obtained by gluing
together three spherocylinders, one for each molecular segment. The Verlet list is then filled 
with the neighboring molecules whose bent spherocylinders 
overlap with the one under consideration.
The bent spherocylinder of a molecule remains fixed for many Monte Carlo steps, but has to be reconstructed
when a rotational or a translational Monte Carlo move brings the molecule out of its 
bent spherocylinder.
Shrinking the spherocylinder radius for the Verlet list considerably reduces the time needed to detect potential overlaps, but increases the rate at which
the lists need to be updated\cite{Frenkel1997}.
Before starting a Monte Carlo simulation, we thus first determine
the spherocylinder radius that optimizes algorithmic performance.

The 
linked list is
built by using a square decomposition of the simulation box. The simulation box is divided into 
smaller sub-boxes of a side 
length 
roughly given by the particle length plus the spherocylinder radius of the Verlet list. 
The Verlet list in a given sub-box can then be built by
only considering molecules within that same sub-box as well as within the eight neighboring sub-boxes.


\subsection{Cluster moves}

In the vicinity of the isotropic--quasi-nematic transition, small clusters of very close and well aligned molecules develop 
in the isotropic phase. Once these relatively small but highly packed clusters of particles form, molecule orientations can 
get kinetically locked. Equilibration, however, requires overcoming the high free-energy barriers associated with aligning 
these clusters
which considerably slows down equilibration.
A way to alleviate this problem relies on the fact that Monte Carlo simulations need not be tied to a local
and thus physical
dynamics. Instead of moving particles one at a time, one can define collective moves that identify groups of correlated particles and then move them as a single object\cite{Wang1990,Dressts1995}.
However, in order to satisfy detailed balance and ensure that 
reverse displacements are also possible, the operation must be done in a probabilistic way. 

Here, we build clusters of particles by introducing an artificial
attractive potential $u_{f}(\epsilon_{ij})$ that links particles together\cite{Whitelam2007,Wu1992}. 
The potential form is formally arbitrary, but is most successful if it captures the nature of correlations within a cluster. Our choice is
\begin{equation}
u_{f}(\epsilon_{ij})=-\epsilon\mathrm{H}(\Delta\theta_{\mathrm{max}}-\Delta\theta_{ij})\mathrm{H}(\Delta r_{\mathrm{max}}-\Delta r_{ij}) \, ,
\end{equation}
where $\mathrm{H}$ is the Heaviside step function, and $\Delta r_{\mathrm{max}}$ and $\Delta\theta_{\mathrm{max}}$ are tunable thresholds. Typical values are $\epsilon=1.0$, $\Delta r_{\mathrm{max}}=0.3$ and $\Delta\theta_{\mathrm{max}}=6.0^{\circ}$.
For a given configuration, the linking probability between particles $i$ and $j$ is
\begin{equation}
p_{ij}= \mathrm{max}[0,1-\exp(\beta u_{f}(\epsilon_{ij}))] \, .
\end{equation}
Molecules that are close and well aligned are thus linked with high probability, 
whereas molecules outside the range of $\Delta r_{\mathrm{max}}$ and $\Delta\theta_{\mathrm{max}}$ are not linked at all.
A cluster is built by choosing a particle $i$ at random and by attempting to build links with its neighbors. If a link between 
$i$ and $j$ is formed, $j$ becomes a member of the cluster formed around molecule $i$. 
Once the cluster has been formed, we perform a trial move by translating 
or by rotating it around its center of mass. If the trial move does not generate any overlap with other molecules,
it is accepted with probability
\begin{equation}
W(\mathrm{old} \rightarrow \mathrm{new}) = \mathrm{min}[1,\exp(\beta(U_{\mathrm{new}}-U_{\mathrm{old}}))] \, ,
\end{equation}
where $U_{\nu}=\sum_{i,j\in I_{\nu}}u_{f}(\epsilon_{ij})$ is the artificial interaction energy between the cluster and its 
environment in either the $\nu=$old or new state.
The interface $I_{\nu}$ is defined by all the particles $i$ inside and the particles $j$ outside the cluster that contribute
to $U_{\nu}$.


\subsection{\label{subsect: excluded area} Evaluation of excluded areas}

In Sec.\ \ref{sec: Onsager Theory} we use Onsager theory \cite{Onsager1949,Martinez-Raton2005} 
as an alternative method for calculating
phase diagrams. This approach heavily depends on the the concept of excluded area.
\begin{figure}
\begin{center}
\includegraphics{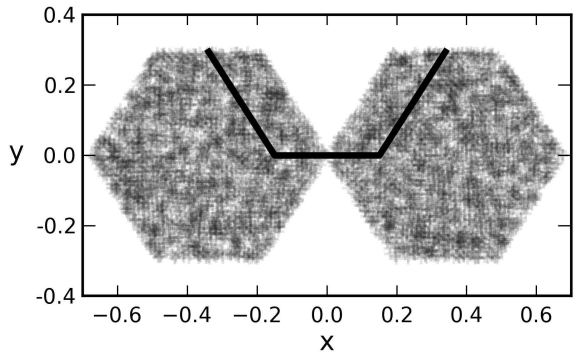}
\caption{Illustration of the calculation of the excluded area for bow-shaped molecules 
with $ \alpha= 57.3 ^{\circ}$, $a=0.35$ and the same orientation.
The excluded area is obtained by repeatedly inserting the
second molecule inside the box and checking for overlaps. 
The shaded region is obtained by placing a dot at the position of the center of mass of a randomly inserted particle 
if it overlaps with the fixed particle.}
\label{fig: excluded area}
\end{center}
\end{figure}
Given two molecules with a 
fixed relative orientation 
$\Delta \theta$, then
the excluded area 
$A_{\mathrm{exl}}(\Delta \theta)$
is defined as the portion of the plane surrounding a molecule that is
not accessible to the other. If the center of the second molecule 
is inside the area excluded by the first, then molecules 
overlap. Figure\ \ref{fig: excluded area} shows an example of the shape of the excluded area for two bow-shaped molecules with the same orientation 
($\Delta \theta = 0$). For hard-core potentials the excluded area 
is proportional to the second virial coefficient
in an expansion of the free energy in powers of density.
Hence, $A_{\mathrm{exl}}$ provides a microscopic description of the 
balance between positional and orientational entropy.

We evaluate the excluded area by Monte Carlo integration.
One molecule is fixed at the center of a box of area $A_{\mathrm{box}}$. The size of the box is chosen 
such that it is impossible for a molecule outside the box to overlap with the fixed molecule. Another molecule with fixed 
relative orientation
$\Delta \theta$ is then inserted at random in the box. The process is repeated $N_{\mathrm{trial}}$ times, keeping track of the number of overlaps $N_{\mathrm{overlap}}$ that occurred during the whole process.
In the end, the excluded area is $A_{\mathrm{exl}}(\Delta \theta) = A_{\mathrm{box}} (N_{\mathrm{overlap}} / N_{\mathrm{trial}})$.


\section{Quasi-Long-Range Orientational Order}
\label{sec.quasi}

According to the Mermin-Wagner theorem the spontaneous symmetry breaking of a continuous order parameter is always 
suppressed by fluctuations in dimensions $d\le2$ for systems with sufficiently short-ranged interactions \cite{Merminf1966}.
Nevertheless, as first pointed out by Kosterlitz and Thouless (KT), a phase transition in $d \le 2$ is still possible between, on the one hand,
a disordered phase wherein the correlation function of the order parameter decays exponentially and thereby only exhibits 
short-range order, and, on the other hand, a phase with quasi-long-range order wherein the correlation function decays as a 
power law in distance $r$ \cite{Kosterlitz1973}.

Since it will be important for our analysis in Sec.\ \ref{subsec: nematic}, we
summarize the main KT results. The description starts from
the free energy associated with distortions 
of a molecular orientational field, 
\begin{equation}
F=\frac{K}{2}\int [\nabla \theta (\vec{r})]^{2}\,
\text{d}^2r  \, ,
\label{eq: frank elastic energy}
\end{equation}
where the angle $\theta(\vec{r})$ measures orientation of the molecule
with respect to a fixed  axis, and $K$ is the 
Frank elastic constant. In a more general theory, Eq.\ (\ref{eq: frank elastic energy}) should include two different elastic constants, 
one for splay and one for bend deformations, but on sufficiently large length scales they  
renormalize
to the same value\cite{Nelson1977}.
The orientational correlation function 
$g_{2}(r)$ introduced in Eq.\ (\ref{eq: orientational correlation}) quantifies the observed orientational order. 
According to Kosterlitz and Thouless,
quasi-long-range orientational order (here, a quasi-nematic phase), results from 
the  competition between the free energy needed 
to create
topological defects and the entropy gained when these defects unbind and  are thus free to move\cite{chaikin2000principles}.
Disclination unbinding takes place at the critical value of the Frank elastic constant\cite{Bates2000},
\begin{equation}
\frac{\pi K_{c}}{8 k_{\text{B}} T} = 1 \, .
\label{eq: critical K}
\end{equation}
We stress that $K_{c}$ is a scale-free quantity that
locates the transition between short-range and quasi-long-range nematic order in the thermodynamic limit \cite{Bates2000}.
The critical value is the result of a balance between the disclination energy and entropy, and in two dimensions both have the 
same logarithmic dependence on system size. If $K < K_{c}$, isolated disclinations 
are found, which leads to an overall isotropic state characterized by an exponential decay of the orientational correlation function.
If $K>K_{c}$, 
disclinations are 
bound in pairs
and the orientational correlation function is thus expected to decay algebraically, $g_{2}(r)\propto r^{-\eta }$, with an exponent \cite{Kosterlitz1973}
\begin{equation}
\eta = 2 k_{\text{B}} T / \pi K \, .
\label{eq: eta vs K}
\end{equation}

In the limit $\alpha=0$, (or for $a=0$) both configurations of the bent-needle model reduce to a straight needle, which is known to
undergo an isotropic--quasi-nematic transition via 
disclination unbinding 
\cite{Frenkel1985,Vink2009}. We demonstrate in Sec.\ \ref{subsec: nematic} that the isotropic--quasi-nematic transition in the bent-needle model is also consistent with 
the KT picture.


\section{\label{sec: result}Results}
In this section we present the results of the Monte Carlo simulations 
described in 
Sec.\ \ref{sec: Model}. We 
detail
how we identify the
isotropic--quasi-nematic transition 
as well as the subsequent quasi-nematic--smectic transition. We also introduce the modulated-nematic phase and
summarize our results in phase diagrams.

\subsection{\label{subsec: nematic}Isotropic--quasi-nematic transition}

\begin{figure}
\includegraphics{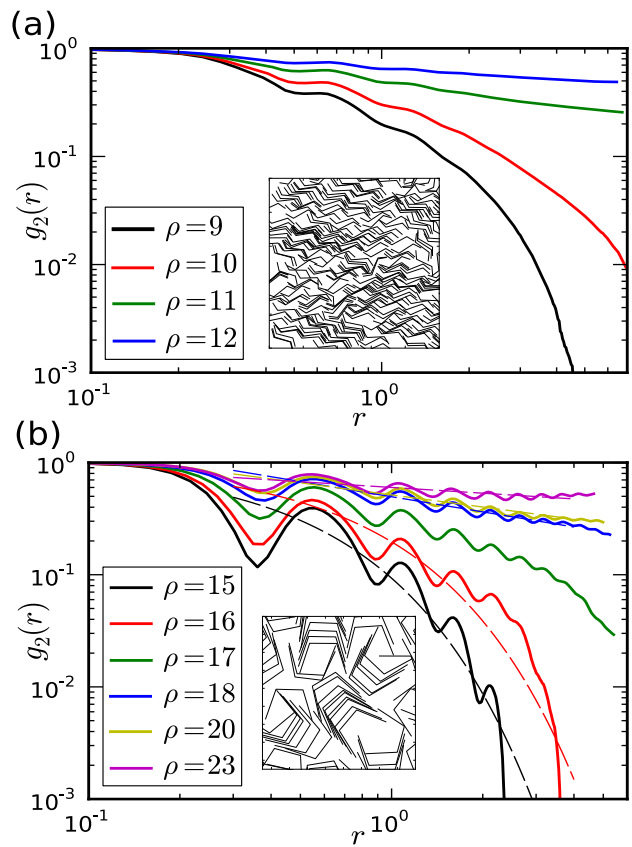}
\caption{(colors online) (a) Radial dependence of the orientational correlation function 
for several densities for
(a) zig-zag molecules 
with $\alpha=\pi/3$ and $a=0.25$, and (b) bow-shaped  molecules with $\alpha=2\pi/5$ and $a=0.35$.
Dashed lines in (b) show the fits to an exponential function for $\rho=15$ and 16 and to $g_2(r)\propto r^{-\eta}$ for $\rho=18$, 20, and 23. Fit exponents are given in Fig.\ \ref{fig: Frank Critical Density}.
The inset in (a) shows a portion of the simulation box at $\rho=11$, close to the isotropic--quasi-nematic transition. The inset in (b) details the local packing of molecules, which results in oscillations
of $g_2(r)$.
}
\label{fig: Correlation Function}
\end{figure}

Upon increasing density, both zig-zag and bow-shaped molecules form a quasi-nematic phase with quasi-long-range orientational order .
The quasi-nematic phase and the isotropic--quasi-nematic phase transition via disclination 
unbinding are evidenced by the transition from an exponential decay of $g_{2}(r)$ at low $\rho$ to
a power-law decay, $g_2(r) \propto r^{-\eta}$, as $\rho$ increases (see Fig.\ \ref{fig: Correlation Function}). 
As per the discussion in Sec.\ \ref{sec.quasi}, the quasi-nematic phase
is expected to be stable against spontaneous disclination unbinding when $\pi K/( 8 k_{\textit{B}}T)>1$.
We obtain the
Frank elastic constants  from the relation $\eta=2 k_{\text{B}} T / \pi K$, where the power-law decay of $g_{2}(r)$ 
is obtained from a linear fit of $ \ln g_{2}(r)$  versus $\ln r$ over the range $0.5 \le r \le 4.0$. Note that the lower threshold is necessary because
$g_{2}(r)$ deviates from the power law at small $r$, while the higher threshold is chosen so as to exclude
correlations
resulting from the use of periodic boundary conditions.
Note also that bow-shaped molecules display correlation functions with fairly large oscillations as a result of the local 
packing structures, which do not exist for straight needles [see inset in Fig.\ \ref{fig: Correlation Function}(b)].

\begin{figure}
\includegraphics{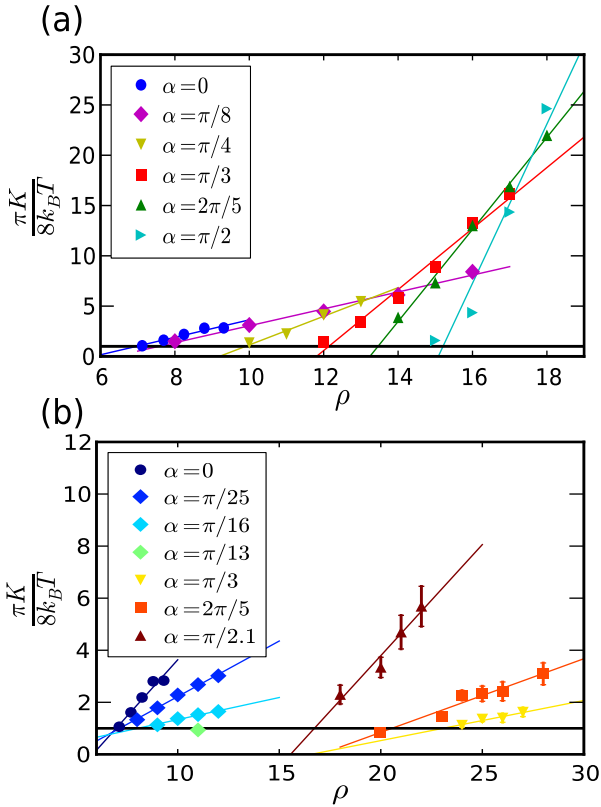}
\caption{(colors online) 
Density evolution of the reduced Frank elastic constant $K$ 
obtained from the power-law decay of $g_2(r)$
for (a) zig-zag molecules with $a=0.25$ and (b) bow-shaped molecules with $a=0.35$. 
The isotropic--quasi-nematic transition densities, $\rho_{\mathrm{IN}}$ are obtained from the intersections
of the linear fits
to the numerical results with the line $\pi K / 8 k_{\mathrm{B}}T = 1$. Where not shown, errorbars are smaller than the symbols. 
}
\label{fig: Frank Critical Density}
\end{figure}

The results for the reduced Frank elastic constant $\pi K/(8k_{\textit{B}}T)$ are plotted
in  Fig.~\ref{fig: Frank Critical Density} for both molecule types. 
Since the results scale nearly linearly with $\rho$, 
we identify the isotropic--quasi-nematic transition densities $\rho_{\mathrm{IN}}$ as the intersection between linear fits
to data the points
and $\pi K_c/(8k_{\textit{B}}T)=1$\cite{Bates2000}.

For zig-zag molecules, the quasi-nematic phase is systematically destabilized by bending the terminal segments 
[Fig. \ref{fig: Frank Critical Density}(a)].
As the central- to-tail angle $\alpha$ increases, the transition is thus pushed to higher densities.
Bow-shaped molecules, however, 
show a non-monotonic 
trend
of the transition density $\rho_{\mathrm{IN}}$ with $\alpha$ [Fig. \ref{fig: Frank Critical Density}(b)].
For small $\alpha$ the transition density increases with $\alpha$, 
while for $\alpha \ge \pi/3$ the trend is inverted. 
In the range $\pi/13\lesssim\alpha\lesssim\pi/3$ the quasi-nematic phase is unstable with respect to the modulated-nematic 
phase, and the power-law scaling analysis is then inapplicable (see Secs.\ \ref{subsec: supra layering}
and \ref{sec.phase}).

\begin{figure}
\begin{center}
\includegraphics{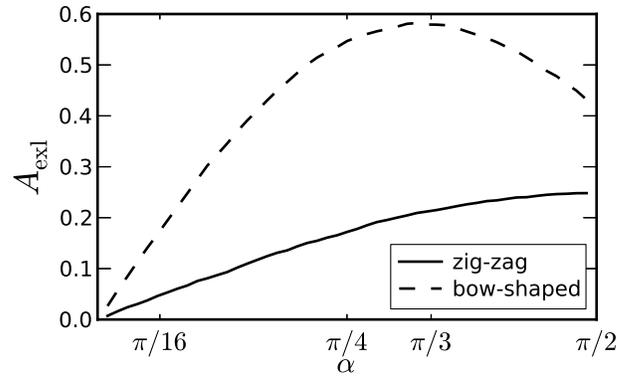}
\caption{Excluded area versus central-to-tail angle $\alpha$ for parallel zig-zag (continuous line) and bow-shaped (dashed line) molecules.}
\label{fig: parallel excluded area}
\end{center}
\end{figure}

We try to understand the difference in the two molecular geometries by examining the size $A_{\mathrm{exl}}$
of the excluded area for  perfectly parallel molecules (see Fig. \ref{fig: parallel excluded area}).
Like $\rho_{\mathrm{IN}}$, it monotonically increases with $\alpha $ for zig-zag molecules, but reaches a maximum 
at $\alpha \approx \pi/3$ and then decreases for bow-shaped molecules.
Large values of $A_{\mathrm{exl}}$ suggest that parallel molecules have to pack locally to come close to one another,
as evidenced by the undulations of $g_2(r)$ in Fig.\ \ref{fig: Correlation Function}(b). This packing constraints 
translational freedom, and thus reduces the translational contribution to the entropy. Because the translational contribution normally compensates for the loss of orientational freedom in the nematic phase, higher $\rho$ than usual are needed for this effect to be significant, and as a result $\rho_{\mathrm{IN}}$ increases.

\subsection{\label{subsec: smectic}Quasi-nematic--smectic transition}

\begin{figure}
\begin{center}
\includegraphics{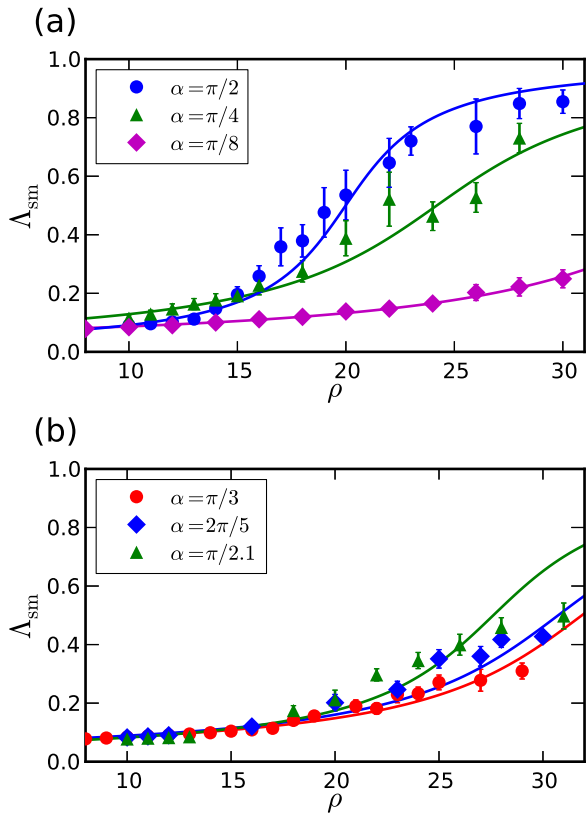}
\caption{(colors online) 
Smectic order parameter $\Lambda_{\mathrm{sm}}$ plotted versus density $\rho$
at several central-to-tail angles $\alpha$
for (a) zig-zag molecules with $a=0.25$ and (b) bow-shaped molecules with $a=0.35$. 
Data points
are obtained from 
Monte Carlo simulations. 
Lines are a fit of Eq.~(\ref{eq:smectic_fit}) to the data
using the transition density $\rho_{\mathrm{NS}}$ and $h$ as
fitting parameters (see phase diagrams of Fig.\ \ref{fig: Phase Diagram} for results on $\rho_{\mathrm{NS}}$).
}
\label{fig: Smectic Order Parameter}
\end{center}
\end{figure}

Smectic order consists of a periodic arrangement of particle positions along one direction, which leads to a well-defined density wave along the corresponding wave vector. Within smectic layers, however, translational order is absent. 
Although long-range translational order is not expected in the thermodynamic limit of two-dimensional systems, 
smectic order is obervable on sufficiently small length scales.
We get back to this point below.
For now, we consider the smectic order parameter defined in Eq.\ (\ref{eq: smectic order parameter}) as a function of 
density for both the zig-zag and bow-shaped molecules in Fig.\ \ref{fig: Smectic Order Parameter}. The Monte Carlo simulation data are fitted with
\begin{equation}
\label{eq:smectic_fit}
f(\rho) = 1/2 + \arctan[h( \rho - \rho_{\mathrm{NS}} )] / \pi,
\end{equation}
where $\rho_{\mathrm{NS}}$ and $h$ are fit parameters. Hence, the quasi-nematic--smectic transition density 
$\rho_{\mathrm{NS}}$ is defined as the point of maximum slope of $f(\rho)$.
The smectic order parameter for bow-shaped molecules does not assume large values even at the 
highest
densities simulated, which is likely the result of out-of-layer fluctuations and
of the instability of two-dimensional smectic
order described. Fits to the simulation results nonetheless provide an estimate for $\rho_{\mathrm{NS}}$. Note that the values of the transition densities identified in this way 
are close to the highest values of the densities explored in our simulation.

As could be seen in Fig.\ \ref{fig: configuration} and was previously noted in Ref.~\onlinecite{Peon2006}, zig-zag 
molecules
form a smectic-C phase, wherein the nematic director is tilted with respect to the layer normal. 
Bow-shaped
molecules instead arrange in an anti-ferromorphic smectic-A structure, wherein the 
polar vector $\hat{\bm{v}}$ adopts an opposite orientation in adjacent layers.
This unusual ordering can be rationalized by extending a packing argument developed for the anti-ferrolectric ordering of V-shaped molecules\cite{Martinez-Gonzalez2012a}, i.e., for $b=0$\cite{Bisi2008}. 
Because the excluded area of two molecules
is smaller in the anti-parallel ($\hat{\bm{v}}^{(i)} = -\hat{\bm{v}}^{(j)}$) than in the parallel 
($\hat{\bm{v}}^{(i)} = \hat{\bm{v}}^{(j)}$) arrangement,
the former is entropically favored.
Molecules can also more easily penetrate into neighboring layers in the anti-ferromorphic 
than in the ferromorphic smectic phase. This effect enhances out-of-layer fluctuations and thus entropically favors the 
anti-ferromorphic smectic ordering
as well.
This behavior is similar to that of three-dimensional bent-core molecules\cite{Lansac2003}.

No significant smectic ordering is found for values of $\alpha$ smaller than those given in 
Figs.\ \ref{fig: Smectic Order Parameter}(a) and (b).
For zig-zag molecules at even smaller $\alpha$,
we extrapolate the quasi-nematic--smectic transition to take place at $\rho_{\mathrm{NS}}$ that are inaccessible within a reasonable computational time; for bow-shaped molecules, however, the existence of a smectic phase at $\alpha < \pi/3$, even for large $\rho$, 
is unclear.

As mentioned above and discussed in detail by Toner and Nelson 
in Ref.~\onlinecite{Toner1981}, two-dimensional long-range smectic order should not be thermodynamically stable.
One expects instead the following scenario. At a given $\rho$ translational order is only disturbed by phonon fluctuations 
in regions with linear dimension smaller than a characteristic length $\xi_{d}$, which is the mean distance between 
thermally induced dislocations. These dislocations thus destroy translational order on length scales larger than $\xi_d$, 
but correlations in the layer orientations persist and exhibit a long-range algebraic decay. Although this scenario is physically reasonable and may explain the 
weak smectic ordering of bow-shaped molecules,
we were unable to test it against our simulation data because of the limited range of computationally accessible $N$. 
Instead, we find smectic order to be fully stabilized in our simulations.


\subsection{\label{subsec: supra layering}Modulated-nematic phase}

\begin{figure}
\begin{center}
\includegraphics{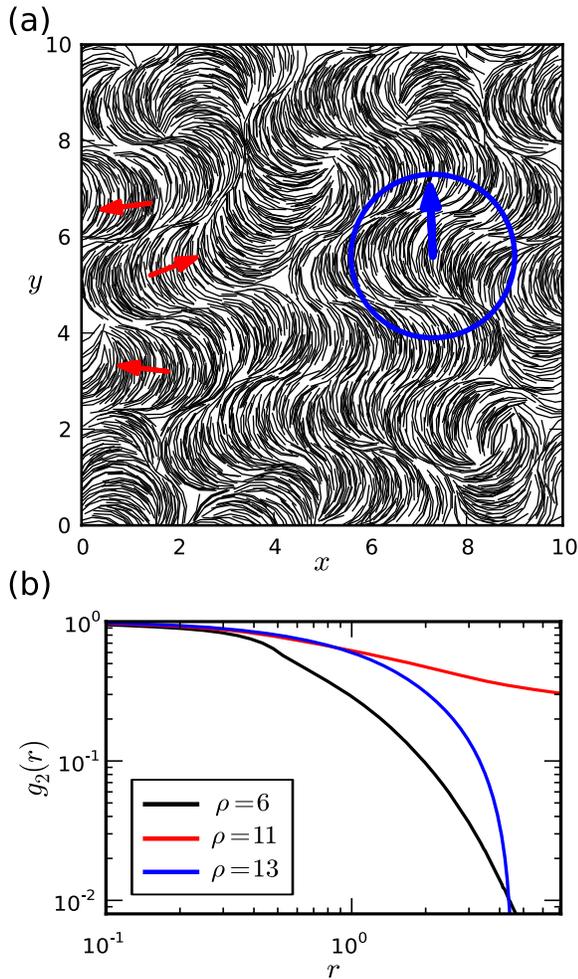}
\caption{(a) Snapshot of bow-shaped molecules in the modulated-nematic phase for
$\alpha=\pi/8$, $a=0.35$, and $\rho=20$. 
The complete simulation box with $N=2000$ molecules is shown.
The red arrows indicate the 
polar vector $\hat{\bm{v}}$ perpendicular to the molecular orientation
in the center of the layers. The blue arrow indicates the direction of the optimal wave vector $\bm{h}_m$.
The radius of the blue circle is half the optimal wavelength $\lambda_m$ as defined in Eq.\ (\ref{eq: modulated order parameter}) 
and obtained from the maximization procedure illustrated in Fig.\ \ref{fig: Fourier Maximization}(b). For this configuration $\Psi_{\mathrm{m}} \approx 0.25$. (b) The orientational correlation function $g_2(r)$ for bow-shaped molecules with $\alpha=\pi/13$ decays exponentially at $\rho=6$, algebraically at $\rho=11$, and exponentially again at $\rho=13$ due to the appearance of the modulated-nematic phase. 
}
\label{fig: supramolecular layering}
\end{center}
\end{figure}

Although at small $\alpha$ and for $\alpha \gtrsim\pi/3$ bow-shaped molecules form a quasi-nematic phase [Fig.\ \ref{fig: configuration}(b)],
for intermediate $\alpha$, they also
equilibrate
in a \emph{modulated-nematic} phase. In this phase no overall orientational or positional order
exists [Fig.\ \ref{fig: supramolecular layering}(a)].
Instead, it shows a different kind of supramolecular arrangement,
wherein the orientation of a series of molecules varies gradually along arches that form approximate half-circles,
and these arches themselves form layers.
Periodic order exists along the layer normal but the mean polar vector $\langle \hat{\bm{v}} \rangle$ in one layer is 
antiparallel  to that of a neighboring layer.
This arrangement destabilizes
the quasi-nematic phase and its
algebraic orientational order.
As shown in Fig. \ref{fig: supramolecular layering}(b), the occurrence of the modulated-nematic phase can even result in a reentrant exponential decay of $g_2(r)$ with increasing $\rho$. 
For $\pi/13 \lesssim \alpha \lesssim \pi/3$, however, we found no evidence for a power-law decay of $g_2(r)$ in the whole range of explored densities
and a direct transition from the isotropic to modulated-nematic phase occurs.

\begin{figure}
\begin{center}
\includegraphics{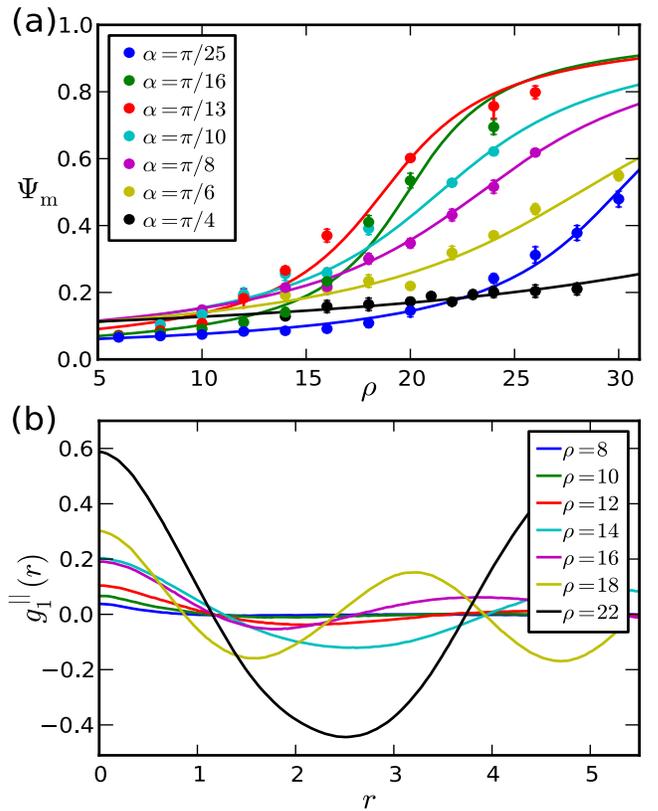}
\caption{(colors online) (a) 
Modulated-nematic order parameter $\Psi_\mathrm{m}$ plotted versus $\rho$ for bow-shaped molecules at several values 
of $\alpha$. The Monte Carlo simulation results are fitted with Eq.\ (\ref{eq: modulated fit}).
From the fit we obtain $\rho_{\mathrm{m}}=29.95$, 19.97, 18.71, 21.42, 23.38, and 45.57 for $\alpha=\pi/25$, $\pi/16$, $\pi/13$, $\pi/10$, $\pi/8$, $\pi/6$, and $\pi/4$, respectively.
(b) Polar correlation function $g_1^{\vert \vert}(r)$ for $\alpha=\pi/10$ and several values of $\rho$. As density increases, the amplitude of the modulation of $g_1^{\vert \vert}(r)$ increases, which we take to be a signature of the modulated-nematic phase.}
\label{fig: modulated-nematic order parameter}
\end{center}
\end{figure}

In Fig.\ \ref{fig: modulated-nematic order parameter}(a) the modulated-nematic order parameter $\Psi_\mathrm{m}$ defined in 
Eq.\ (\ref{eq: modulated order parameter}) is shown as a function of density for several values of $\alpha$.
In analogy with the treatment for the identification of smectic order in  Sec. \ref{subsec: smectic}, we fit $\Psi_\mathrm{m}$ 
(as obtained from the Monte Carlo simulation data) with a trial function similar to Eq.\ (\ref{eq:smectic_fit}),
\begin{equation}
f(\rho)=1/2+\arctan[l(\rho-\rho_{\mathrm{m}})]/\pi
\label{eq: modulated fit}
\end{equation}
where $l$ and $\rho_{\mathrm{m}}$ are fit parameters.
In order to clarify the structural properties of the modulated-nematic phase, we consider the polar correlation 
function $g_1^{\vert \vert}(r)$ as defined in Eq.\ (\ref{eq: modulated-correlation-function}).
Because of the periodic modulation in the molecular polar vector $\hat{\bm{v}}$ along the layer normal, $g_1^{\vert \vert}(r)$ 
becomes a periodic function 
when the layered structure is well established.
In particular, 
since
molecules in adjacent layers have opposite polarization, $g_1^{\vert \vert}(r)$ 
shows a minimum at a distance corresponding to the layer thickness
followed by a maximum, which results from correlations with the next-nearest-neighbor layer.
In Fig.\ \ref{fig: modulated-nematic order parameter}(b) we show $g_1^{\vert \vert}(r)$ for $\alpha=\pi/10$ and several values 
of  $\rho$. As expected, the periodic modulation of $g_1^{\vert \vert}(r)$ becomes stronger with increasing density indicating the progressive development of layers. This behavior is also observed for all other values of $\alpha$. 

\begin{figure}
\begin{center}
\includegraphics{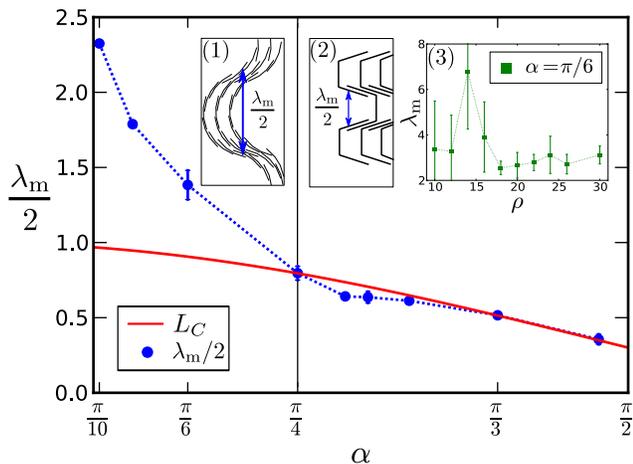}
\caption{(colors online) 
Distance between two adjacent layers, $\lambda_{\mathrm{m}}/2$, 
plotted versus $\alpha$ (points).
The results for $\lambda_{\mathrm{m}}$ are obtained at $\rho>\bar{\rho}_{\mathrm{m}}$, but are observed to be fairly insensitive to changes in $\rho$ [inset (3)].
The continuous red line 
is the projected length of the molecule onto a direction
along the central molecular segment. The vertical line at $\alpha=\pi/4$ approximately divides the plot in two regions. 
The first region with $\alpha<\pi/4$ has $\lambda/2 > L_C$ and molecules are arranged in the supra-molecular structure of the 
modulated-nematic phase illustrated in inset (1). The second region with $\alpha>\pi/4$ has $\lambda/2 \lesssim L_C$, where 
molecules locally arrange with anti-ferromorphic order with adjacent layers slightly inter-penetrated, as illustrated in inset (2).
The dotted line is a guide to the eyes.}
\label{fig: wavelength versus projection}
\end{center}
\end{figure} 

Our Monte Carlo data show that modulations in $g_1^{\vert \vert}(r)$ are already present at relatively small $\rho$, whereas 
the order parameter $\Psi_{\mathrm{m}}$ has not yet increased much. For instance, for $\alpha=\pi/10$ it can be seen in 
Fig.\ \ref{fig: modulated-nematic order parameter}(b) that $g_1^{\vert \vert}(r)$ is already weakly modulated at $\rho=14$, while 
the corresponding order parameter is only $\Psi_{\mathrm{m}} \approx 0.2$. Similarly, we find $\Psi_{\mathrm{m}} \approx 0.25$ 
for the configuration shown in Fig.\ \ref{fig: supramolecular layering}(a), even though a remarkable degree of layering is already 
clearly established. In other words, 
the layered structure of the modulated-nematic phase is established continuously,
which makes it difficult to clearly establish a transition density.  
For this reason, we approximate the transition density $\bar{\rho}_{\mathrm{m}}$ into the modulated-nematic phase by using 
the threshold $\Psi_{\mathrm{m}}=0.2$ on the order parameter. The same threshold is used for all $\alpha$. This particular value is chosen in such a way that $\bar{\rho}_{\mathrm{m}}$ matches reasonably well the values of the density at which we observe the reentrant exponential decay of $g_2(r)$ at $\alpha \lesssim \pi/13$.

The maximization procedure illustrated in Fig.~\ref{fig: Fourier Maximization}(b) for determining the order parameter 
$\Psi_{\mathrm{m}}$ also provides the typical distance between two layers in the modulated-nematic phase, i.e., 
$\lambda_{\mathrm{m}}/2$ defined in Eq.\ (\ref{eq: modulated order parameter}). In Fig.~\ref{fig: wavelength versus projection} 
we show how this distance evolves with $\alpha$ and compare it with the projected length of the 
molecule along the direction parallel to the central molecular segment, $L_C=b+2a \cos(\alpha)$. At small $\alpha$, $\lambda_{\mathrm{m}}>L_C$, which captures the supra-molecular structure of the modulated nematic phase. As $\alpha$ increases, however, $\lambda_{\mathrm{m}}$ decreases, and it becomes comparable to $L_C$ at $\alpha \approx \pi/4 $, indicating that 
molecules locally arrange in anti-ferromorphic order, as demonstrated by the inset (2) of 
Fig.~\ref{fig: wavelength versus projection}, which is then favored over the supra-molecular ordering of the modulated-nematic phase. 
For $\alpha>\pi/4$ the layering distance becomes smaller than $L_C$, indicating that adjacent anti-ferromorphic layers
on average slightly inter-penetrate, as discussed in Sec.\ \ref{subsec: smectic}.

The occurrence of a three-dimensional, spontaneously formed, modulated-nematic phase for banana-shaped mesogens and of the two-dimensional modulated-nematic phase for bow-shaped molecules,
likely results from ``pathological elasticity'' \cite{Dozov2001a},
which has been predicted to be a consequence of the molecular curvature radius.
In the standard Frank elastic theory, splay, bend, and twist elastic constants are indeed assumed to be positive 
in order to ensure a ground state with uniform nematic order. A non-uniform nematic ground state can thus be explained 
as resulting from a negative elastic constant, e.g., the bend constant, which is not forbidden by symmetry. 
Fourth-order terms in the elastic free energy are then needed to stabilize the modulated phase \cite{Dozov2001a}.
Our results thus indicate that there is an upper limit for the molecular curvature, corresponding in our model to 
$\alpha \approx \pi/4$, beyond which the supra-molecular structure of the modulated-nematic phase becomes unstable 
in favor of the anti-ferromorphic arrangement.

We already discussed in Sect.\ \ref{subsec: smectic} that long-range order in two-dimensional smectics is not stable.
In Fig.\ \ref{fig: supramolecular layering}(b) in the center a dislocation in the layering of the supramolecular arches
is visible. This might be an indication how, in analogy to two-dimensional smectics,
also
the layered structure 
is destabilized by the proliferation of dislocations in
sufficiently large systems.

\subsection{Phase diagrams} 

\label{sec.phase}

\begin{figure}
\includegraphics{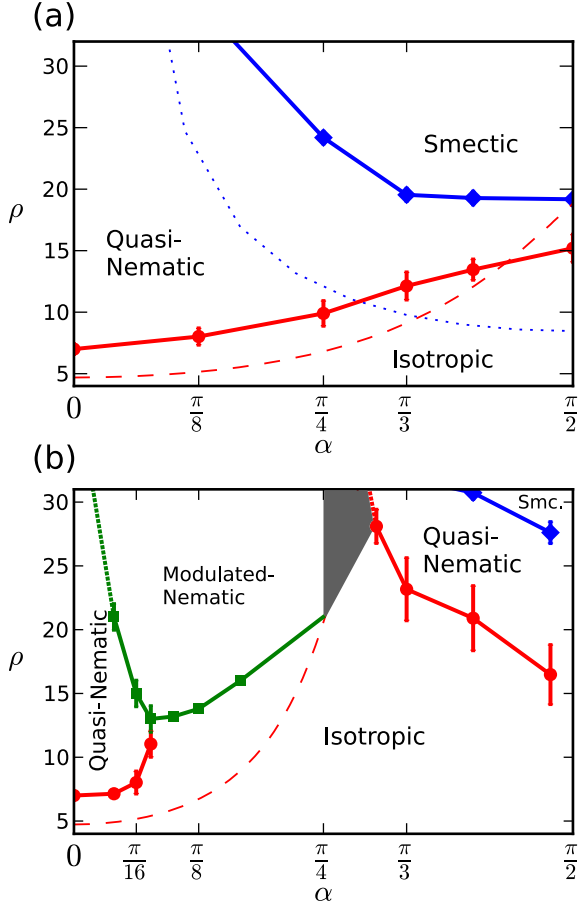}
\caption{(color online) Phase diagram for (a) zig-zag molecules with $a=0.25$ and (b) bow-shaped molecules with $a=0.35$. Points represent the isotropic--quasi-nematic transition densities (circles), quasi-nematic--smectic transition densities (diamonds) and the quasi-nematic-- or isotropic--modulated-nematic transition (squares) as identified from the procedures described in 
Secs.\ \ref{subsec: nematic}, \ref{subsec: smectic} and \ref{subsec: supra layering}, respectively. The gray area in (b) marks the transition between modulated-nematic and quasi-nematic phases in a region where none of the identified mesophases is found to be stabilized. Continuous lines are guides to the eyes. Where not shown, errorbars are smaller than the marker size. Dashed and dotted lines represent, respectively, the isotropic--nematic and the nematic--smectic transition 
lines predicted by Onsager theory (Sec.~\ref{sec: Onsager Theory}).}
\label{fig: Phase Diagram}
\end{figure}

In Fig.~\ref{fig: Phase Diagram} we show the 
simulated
phase diagrams for both the zig-zag and the bow-shaped molecules obtained using the approaches 
described in 
Secs. \ref{subsec: nematic}, \ref{subsec: smectic}, and \ref{subsec: supra layering}.
The dashed red and dotted blue lines show the predictions from Onsager theory  for the transition densities
$\rho_{\mathrm{IN}}$ and $\rho_{\mathrm{NS}}$, respectively.
We will review Onsager theory in Sec.\ \ref{sec: Onsager Theory}.

According to the phase diagram in Fig.\ \ref{fig: Phase Diagram}(a),
terminal segments of zig-zag particles destabilize the quasi-nematic phase but stabilize the smectic phase.
This qualitative trend is captured by Onsager theory. Our simulation results are in rough
agreement with previous studies\cite{Varga2009b,Peon2006} but also exhibit some 
significant discrepancies. In particular, in our phase diagram both the isotropic--quasi-nematic and the 
quasi-nematic--smectic transitions appear at higher $\rho$ (Fig.\ \ref{fig: Phase Diagram}(a)). 
In addition, for $\alpha \approx\pi/2$, the isotropic and smectic phases are well separated by the quasi-nematic phase, 
while the latter was not observed at all in previous studies.
Most likely, the lower transition densities reported in Refs.\ \onlinecite{Varga2009b,Peon2006} follow from 
using the nematic order parameter for detecting the isotropic--quasi-nematic transition
and from considering relatively small system sizes. As already mentioned, the nematic order parameter in a quasi-nematic
phase decreases with system size \cite{Vink2009}.

Bow-shaped molecules exhibit a remarkably rich
behavior.\\
\textbf{(1)} For small $\alpha$, the low-density behavior is similar to that of zig-zag molecules.
The transition density $\rho_{\mathrm{IN}}$ (red circles) increases with increasing $\alpha$ and is
underestimated by Onsager theory. Quasi-nematic order in this small-$\alpha$ region is also
destabilized by further increasing the packing density. Instead of the emergence of a smectic phase, however, a
modulated-nematic phase takes precedence. We did not specifically explore the phase behavior of bow-shaped 
molecules with $\alpha<\pi/25$, but we expect the modulated-nematic phase to appear at ever higher densities as
$\alpha$ decreases, as indicated by the 
dotted green line. For symmetry reasons the bent-nematic phase 
cannot exist for $\alpha=0$.\\
\textbf{(2)} For
$\pi/13\lesssim\alpha\lesssim\pi/4$,
both the quasi-nematic and the
smectic orders are destabilized by 
the modulated-nematic phase at all $\rho$ explored. The direct isotropic--modulated-nematic transition appears at increasing densities with $\alpha$. This phase is simply not captured by 
Onsager theory as formulated in Sect. \ref{sec: Onsager Theory}.\\
\textbf{(3)} For $\pi/4 \lesssim \alpha \lesssim \pi/3$, which is indicated in the phase diagram as a gray area, we find no clear evidence for any of the mesophases known to be formed by bow-shaped molecules. In this region
the distance between two adjacent layers approaches the projection of the total molecular length along the molecular central segment, resulting in a strong competition between the formation of supra-molecular layers and anti-ferromorphic domains. The investigation of configurations at densities much higher than the ones we could reach would be necessary to clarify what mesophase, if any, is stable in that system.   \\
\textbf{(4)} For $\pi/3 \lesssim \alpha < \pi/2$, the molecular curvature is too large to induce 
a spontaneous bending of the nematic director and the corresponding layer formation.
Quasi-nematic order then becomes stable again, but this time the transition densities 
decrease with increasing $\alpha$.
The isotropic--quasi-nematic transition is highly overestimated by Onsager theory.
The theory strongly relies on the excluded area, as discussed in Sec.\ \ref{subsect: excluded area}, but
the excluded area of bow-shaped molecules is minimal in the anti-parallel configuration. In our simulations 
we instead observe clusters of molecules packed in parallel.
Because this arrangement involves structural correlations between more than two molecules, it is not surprising that Onsager theory, which only takes into account two-particle correlations dramatically fails.

Note that we did not study molecules with $\alpha=\pi/2$. 
With this central-to-tail angle
parallel molecules cannot anymore  be shifted into each other to form closely packed clusters.
A different phase is thus expected. Tetradic order has indeed been observed in related models\cite{Martinez-Gonzalez2012}.


\section{\label{sec: Onsager Theory}Onsager Theory for Phase Diagrams}

In 
Ref.~\onlinecite{Varga2009b}, 
a density functional theory 
in mean-field approximation
was used to determine
the phase behavior of zig-zag 
molecules.
Based on the
second-order virial expansion of the 
free energy
introduced by Onsager,
predictions for both $\rho_{\mathrm{IN}}$ and $\rho_{\mathrm{NS}}$ were made~\cite{Varga2009b}. 
In the following, we summarize and extend the results of 
this Onsager theory.

We start with the free energy functional
\begin{eqnarray}
\beta F &= & \int d\vec{r} \int d\vec{\omega} \{ \ln\rho(\vec{r},\vec{\omega})-1\}  \nonumber \\
& &-\frac{1}{2}\int d\vec{r}_1 \int d \vec{\omega}_1 \rho(\vec{r}_1 , \vec{\omega}_1) \nonumber \\
& & \times \int d\vec{r}_2 \int d \vec{\omega}_2  \rho(\vec{r}_2 , \vec{\omega}_2) f_{M}(\vec{r}_{12},\vec{\omega}_1,\vec{\omega}_2) \, .
\label{eq: Onsager free energy}
\end{eqnarray}
where $\rho(\vec{r},\vec{\omega})$ is the local number density in terms of particle
position $\vec{r}$ and particle orientation  $\vec{\omega}$. The Mayer function $f_M$ of the pair potential
is simple for hard-core particles. It is zero except when particles overlap, where $f_M=-1$.

In the isotropic and nematic phases, positional order does not exist and one has
$\rho(\vec{r},\vec{\omega})=\rho f(\vec{\omega})$.
In the second term of the Onsager free energy functional\ (\ref{eq: Onsager free energy}), this introduces the excluded area 
$A_{\mathrm{exl}} (\alpha)$ in terms of the angle $\alpha$ between the two molecular orientations in two dimensions.
Expanding orientational distribution $f(\vec{\omega}) = f(\theta)$ and $A_{\mathrm{exl}} (\alpha)$ into Fourier modes and
minimizing with respect to the Fourier amplitudes of $f(\theta)$, gives a series of possible
bifurcation densities
\begin{equation}
\rho_{B}^{(n)} = -\frac{2}{A_{n}} \, ,
\label{eq: rho I-N}
\end{equation}
Here,
$A_{n}$ is the $n$-th Fourier amplitude of the
excluded area $A_{\mathrm{exl}}(\alpha)$,
\begin{equation}
A_{n}=\frac{2 m}{\pi} \int \limits_{0}^{\pi/m} A_{\mathrm{exl}}(\alpha) \cos( m n \alpha) \, d \alpha \, .
\label{eq: excluded area fourier component}
\end{equation} 
We have introduced an index $m=2$ for zig-zag molecules and $m=1$ for bow-shaped molecules
because their respective excluded areas either have a period $\pi$ or $2\pi$.
We determine the Fourier coefficients $A_n$
numerically by means of the Monte Carlo integration technique 
described in
Sec.\ \ref{subsect: excluded area}. 
We choose the lowest positive value of 
$\rho_{B}^{(n)}$
as the transition density $\rho_{\mathrm{IN}}$, which 
is realized at $n=1$ for zig-zag molecules and at $n=2$ for bow-shaped molecules.
The resulting transition lines are plotted as red dashed lines in the phase diagrams of 
Fig.\ \ref{fig: Phase Diagram}.

To determine the
nematic--smectic transition density $\rho_{\mathrm{NS}}$,
we choose a periodic modulation for the density along the $z$ axis, $\rho(\vec{r})=\rho(z)=\rho(z+d)$,
and assume perfect alignment of the molecules as in Ref.\ \onlinecite{Varga2009b}, because the nematic order in the smectic phase is
typically very high.
The evaluation of the free energy functional now involves an excluded distance $d_{\mathrm{exl}}(z,\theta)$, where $\theta$ is the orientation angle of the central molecular segment with respect to
the smectic layer normal. The excluded distance is related to the excluded area by $A_{\mathrm{exl}} = 
\int d z \,  d_{\mathrm{exl}}(z,\theta) $. Taking into account only the first Fourier mode of the density modulation
in the free energy and minimizing with respect to the Fourier amplitude, we obtain an equation for the
nematic--smectic transition density $\rho_{\mathrm{NS}}$,
\begin{equation}
1+\rho_{\mathrm{NS}} \int \cos(2 \pi z / d) d_{\mathrm{exl}}(z,\theta)\, dz=0 \, ,
\label{eq: smectic transition}
\end{equation}
where  $d$ is the smectic period. The tilt angle $\theta$ and the period $d$ are also determined by minimizing the 
free energy functional.
For
details of the calculation,
we
refer the 
reader to the work of \citeauthor{Varga2009b}\cite{Varga2009b}.
Note that we were only able to calculate a nematic--smectic transition line for zig-zag molecules.
It is plotted in the phase diagram of Fig.\ \ref{fig: Phase Diagram} (a) as blue dotted line.

As already discussed, for the zig-zag molecules Onsager theory qualitatively captures the behavior of both the 
isotropic--nematic and the nematic--smectic transition lines. Both lines are underestimated by the theory, but that is hardly surprising. Onsager theory relies on a second-order virial expansion of the system's free energy,  which is strictly valid only in the 
dilute regime, while in our model orientational and translational order occur at relatively high densities. 
Indeed, it was shown that virial coefficients of order higher than two are not negligible in hard-needle systems \cite{Frenkel1985}. 
In addition, Onsager theory does not take into account topological defects that play a significant role in the two-dimensional phase
behavior and phase transitions. In Ref. \onlinecite{Varga2009b} good agreement between predictions from Onsager theory and 
results from Monte Carlo simulations are reported for zig-zag molecules. We believe this agreement to be due in part to 
the small system size of 200 molecules, where defects cannot fully develop.


\section{\label{sec: conclusion}Conclusion}

Despite its simplicity, the bent-needle model discussed in this paper shows a variety of liquid-crystal phases and 
gives an example of how molecular geometry controls their formation. 

Chiral zig-zag molecules assume quasi-nematic and smectic phases, 
depending on density and the central-to-tail angle $\alpha$.
We use
the orientational correlation function $g_2(r)$, which decays exponentially in the isotropic phase and algebraically in the orientationally ordered phase, 
to identify
the isotropic--quasi-nematic transition.
In two-dimensional systems
with 
short-ranged interactions,
one expects the transition to take place via disclination unbinding
\cite{Kosterlitz1973}. 
Indeed, apart from packing effects, the correlation functions described in Sec. \ref{subsec: nematic} 
behave similarly to the ones already reported
in other two-dimensional anisotropic models with hard-core interactions \cite{Bates2000,Frenkel1985,Vink2009,Armas-Perez2011}. 
For the zig-zag molecule the isotropic--nematic transition density $\rho_{\mathrm{IN}}$ increases with increasing 
central-to-tail angle $\alpha$, while the nematic--smectic transition, $\rho_{\mathrm{NS}}$, exhibits the opposite trend. 
Such behavior is qualitatively captured by Onsager theory 
although
both $\rho_{\mathrm{IN}}$ and $\rho_{\mathrm{NS}}$ are underestimated by the theory,
as discussed in Sec.\ \ref{sec: Onsager Theory}.

Furthermore, we observe a smectic\ C phase, where the central segment of the zig-zag molecule is tilted against the
layer normal. The clear formation of smectic layers, however, indicates that our systems are too small to observe the dislocation unbinding scenario
predicted in Ref.\ \onlinecite{Toner1981}.

Achiral bow-shaped molecules have a much richer 
phase behavior.
It can be divided into three regions, depending on the value 
of $\alpha$. For small $\alpha$ 
molecules 
form 
isotropic, quasi-nematic, and modulated-nematic phases, as density increases.
The
isotropic--quasi-nematic transition is defect driven, which 
makes
the orientational correlation function 
switch from an
exponential 
to power-law decay with increasing density. 
No polar 
order
is found in the quasi-nematic phase. 
Further increasing density
destabilizes the orientational order
of the quasi-nematic phase.
A modulated-nematic phase then takes over, wherein bow-shaped molecules form layers of supramolecular arches.
The orientational correlation function $g_2(r)$ shows a reentrant exponential decay that corresponds to the development 
of the supramolecular arches. At intermediate $\alpha$,
a direct transition from the isotropic to the modulated-nematic phase takes place. 
The modulated-nematic structure becomes less pronounced with increasing 
$\alpha$, i.e., the layer thickness decreases towards the molecule length, up to the point where the formation of supramolecular arches becomes unfavored and anti-ferromorphic domains develop instead.
For $\alpha\gtrsim\pi/3$,
the curvature radius of the molecules becomes too small to induce spontaneous bending of the nematic director 
and quasi-nematic order reenters. Anti-ferromorphic smectic order is then found at even higher density.

A very appealing result of our investigation is the identification of the modulated-nematic phase made from supramolecular layers,
which strongly depend on molecular geometry. Our simulations are the first to clearly demonstrate such a layered 
structure in two dimensions and to relate it to molecular geometry. We find the optimal value of the curvature radius to be around $\alpha \approx \pi/10$
when the modulated-nematic phase occurs at the lowest packing density. This finding might be particularly useful in developing novel
functional optical materials based on organic bent-core liquid crystals\cite{etxebarria2008bent,pintre2010bent}, where the formation of polar domains can be used to tune the nonlinear optical properties of the material.

Our investigations further illustrate the richness of structures, including supra-molecular organization, formed by self-assembling particles of different shapes. Being able to control molecular geometry thus offers the possibility of designing 
novel materials, in particular in two dimensions, and of tuning their properties accordingly.

\begin{acknowledgments}

This work was supported by the Deutsche Forschungsgemeinschaft through the international research training group IRTG 1524. PC is thankful for support from the National Science Foundation Research Triangle Materials Research Science and Engineering Center (DMR-1121107).
\end{acknowledgments}




\begin{thebibliography}{76}%
\makeatletter
\providecommand \@ifxundefined [1]{%
 \@ifx{#1\undefined}
}%
\providecommand \@ifnum [1]{%
 \ifnum #1\expandafter \@firstoftwo
 \else \expandafter \@secondoftwo
 \fi
}%
\providecommand \@ifx [1]{%
 \ifx #1\expandafter \@firstoftwo
 \else \expandafter \@secondoftwo
 \fi
}%
\providecommand \natexlab [1]{#1}%
\providecommand \enquote  [1]{``#1''}%
\providecommand \bibnamefont  [1]{#1}%
\providecommand \bibfnamefont [1]{#1}%
\providecommand \citenamefont [1]{#1}%
\providecommand \href@noop [0]{\@secondoftwo}%
\providecommand \href [0]{\begingroup \@sanitize@url \@href}%
\providecommand \@href[1]{\@@startlink{#1}\@@href}%
\providecommand \@@href[1]{\endgroup#1\@@endlink}%
\providecommand \@sanitize@url [0]{\catcode `\\12\catcode `\$12\catcode
  `\&12\catcode `\#12\catcode `\^12\catcode `\_12\catcode `\%12\relax}%
\providecommand \@@startlink[1]{}%
\providecommand \@@endlink[0]{}%
\providecommand \url  [0]{\begingroup\@sanitize@url \@url }%
\providecommand \@url [1]{\endgroup\@href {#1}{\urlprefix }}%
\providecommand \urlprefix  [0]{URL }%
\providecommand \Eprint [0]{\href }%
\providecommand \doibase [0]{http://dx.doi.org/}%
\providecommand \selectlanguage [0]{\@gobble}%
\providecommand \bibinfo  [0]{\@secondoftwo}%
\providecommand \bibfield  [0]{\@secondoftwo}%
\providecommand \translation [1]{[#1]}%
\providecommand \BibitemOpen [0]{}%
\providecommand \bibitemStop [0]{}%
\providecommand \bibitemNoStop [0]{.\EOS\space}%
\providecommand \EOS [0]{\spacefactor3000\relax}%
\providecommand \BibitemShut  [1]{\csname bibitem#1\endcsname}%
\let\auto@bib@innerbib\@empty
\bibitem [{\citenamefont {Ulman}(2013)}]{ulman2013introduction}%
  \BibitemOpen
  \bibfield  {author} {\bibinfo {author} {\bibfnamefont {A.}~\bibnamefont
  {Ulman}},\ }\href@noop {} {\emph {\bibinfo {title} {An Introduction to
  Ultrathin Organic Films: From Langmuir--Blodgett to Self--Assembly}}}\
  (\bibinfo  {publisher} {Academic press},\ \bibinfo {year} {2013})\BibitemShut
  {NoStop}%
\bibitem [{\citenamefont {Schreiber}(2004)}]{schreiber2004self}%
  \BibitemOpen
  \bibfield  {author} {\bibinfo {author} {\bibfnamefont {F.}~\bibnamefont
  {Schreiber}},\ }\href@noop {} {\bibfield  {journal} {\bibinfo  {journal} {J.
  of Phys. Condens. Matter}\ }\textbf {\bibinfo {volume} {16}},\ \bibinfo
  {pages} {R881} (\bibinfo {year} {2004})}\BibitemShut {NoStop}%
\bibitem [{\citenamefont {B{\"o}ker}\ \emph {et~al.}(2007)\citenamefont
  {B{\"o}ker}, \citenamefont {He}, \citenamefont {Emrick},\ and\ \citenamefont
  {Russell}}]{boker2007self}%
  \BibitemOpen
  \bibfield  {author} {\bibinfo {author} {\bibfnamefont {A.}~\bibnamefont
  {B{\"o}ker}}, \bibinfo {author} {\bibfnamefont {J.}~\bibnamefont {He}},
  \bibinfo {author} {\bibfnamefont {T.}~\bibnamefont {Emrick}}, \ and\ \bibinfo
  {author} {\bibfnamefont {T.~P.}\ \bibnamefont {Russell}},\ }\href@noop {}
  {\bibfield  {journal} {\bibinfo  {journal} {Soft Matter}\ }\textbf {\bibinfo
  {volume} {3}},\ \bibinfo {pages} {1231} (\bibinfo {year} {2007})}\BibitemShut
  {NoStop}%
\bibitem [{\citenamefont {Barth}, \citenamefont {Costantini},\ and\
  \citenamefont {Kern}(2005)}]{barth2005engineering}%
  \BibitemOpen
  \bibfield  {author} {\bibinfo {author} {\bibfnamefont {J.~V.}\ \bibnamefont
  {Barth}}, \bibinfo {author} {\bibfnamefont {G.}~\bibnamefont {Costantini}}, \
  and\ \bibinfo {author} {\bibfnamefont {K.}~\bibnamefont {Kern}},\ }\href@noop
  {} {\bibfield  {journal} {\bibinfo  {journal} {Nature}\ }\textbf {\bibinfo
  {volume} {437}},\ \bibinfo {pages} {671} (\bibinfo {year}
  {2005})}\BibitemShut {NoStop}%
\bibitem [{\citenamefont {Aswal}\ \emph {et~al.}(2006)\citenamefont {Aswal},
  \citenamefont {Lenfant}, \citenamefont {Guerin}, \citenamefont {Yakhmi},\
  and\ \citenamefont {Vuillaume}}]{aswal2006self}%
  \BibitemOpen
  \bibfield  {author} {\bibinfo {author} {\bibfnamefont {D.}~\bibnamefont
  {Aswal}}, \bibinfo {author} {\bibfnamefont {S.}~\bibnamefont {Lenfant}},
  \bibinfo {author} {\bibfnamefont {D.}~\bibnamefont {Guerin}}, \bibinfo
  {author} {\bibfnamefont {J.}~\bibnamefont {Yakhmi}}, \ and\ \bibinfo {author}
  {\bibfnamefont {D.}~\bibnamefont {Vuillaume}},\ }\href@noop {} {\bibfield
  {journal} {\bibinfo  {journal} {Anal. Chim. Acta}\ }\textbf {\bibinfo
  {volume} {568}},\ \bibinfo {pages} {84} (\bibinfo {year} {2006})}\BibitemShut
  {NoStop}%
\bibitem [{\citenamefont {Hicks}\ and\ \citenamefont
  {Petralli-Mallow}(1999)}]{hicks1999nonlinear}%
  \BibitemOpen
  \bibfield  {author} {\bibinfo {author} {\bibfnamefont {J.}~\bibnamefont
  {Hicks}}\ and\ \bibinfo {author} {\bibfnamefont {T.}~\bibnamefont
  {Petralli-Mallow}},\ }\href@noop {} {\bibfield  {journal} {\bibinfo
  {journal} {Appl. Phys. B}\ }\textbf {\bibinfo {volume} {68}},\ \bibinfo
  {pages} {589} (\bibinfo {year} {1999})}\BibitemShut {NoStop}%
\bibitem [{\citenamefont {Mendes}(2008)}]{mendes2008stimuli}%
  \BibitemOpen
  \bibfield  {author} {\bibinfo {author} {\bibfnamefont {P.~M.}\ \bibnamefont
  {Mendes}},\ }\href@noop {} {\bibfield  {journal} {\bibinfo  {journal} {Chem.
  Soc. Rev.}\ }\textbf {\bibinfo {volume} {37}},\ \bibinfo {pages} {2512}
  (\bibinfo {year} {2008})}\BibitemShut {NoStop}%
\bibitem [{\citenamefont {Hore}\ and\ \citenamefont
  {Composto}(2010)}]{hore2010nanorod}%
  \BibitemOpen
  \bibfield  {author} {\bibinfo {author} {\bibfnamefont {M.~J.}\ \bibnamefont
  {Hore}}\ and\ \bibinfo {author} {\bibfnamefont {R.~J.}\ \bibnamefont
  {Composto}},\ }\href@noop {} {\bibfield  {journal} {\bibinfo  {journal} {ACS
  nano}\ }\textbf {\bibinfo {volume} {4}},\ \bibinfo {pages} {6941} (\bibinfo
  {year} {2010})}\BibitemShut {NoStop}%
\bibitem [{\citenamefont {Mclean}\ \emph {et~al.}(2006)\citenamefont {Mclean},
  \citenamefont {Huang}, \citenamefont {Khripin}, \citenamefont {Jagota},\ and\
  \citenamefont {Zheng}}]{mclean2006controlled}%
  \BibitemOpen
  \bibfield  {author} {\bibinfo {author} {\bibfnamefont {R.~S.}\ \bibnamefont
  {Mclean}}, \bibinfo {author} {\bibfnamefont {X.}~\bibnamefont {Huang}},
  \bibinfo {author} {\bibfnamefont {C.}~\bibnamefont {Khripin}}, \bibinfo
  {author} {\bibfnamefont {A.}~\bibnamefont {Jagota}}, \ and\ \bibinfo {author}
  {\bibfnamefont {M.}~\bibnamefont {Zheng}},\ }\href@noop {} {\bibfield
  {journal} {\bibinfo  {journal} {Nano Lett.}\ }\textbf {\bibinfo {volume}
  {6}},\ \bibinfo {pages} {55} (\bibinfo {year} {2006})}\BibitemShut {NoStop}%
\bibitem [{\citenamefont {Slyusarenko}, \citenamefont {Constantin},\ and\
  \citenamefont {Davidson}(2014)}]{slyusarenko2014two}%
  \BibitemOpen
  \bibfield  {author} {\bibinfo {author} {\bibfnamefont {K.}~\bibnamefont
  {Slyusarenko}}, \bibinfo {author} {\bibfnamefont {D.}~\bibnamefont
  {Constantin}}, \ and\ \bibinfo {author} {\bibfnamefont {P.}~\bibnamefont
  {Davidson}},\ }\href@noop {} {\bibfield  {journal} {\bibinfo  {journal} {J.
  Chem. Phys.}\ }\textbf {\bibinfo {volume} {140}},\ \bibinfo {pages} {104904}
  (\bibinfo {year} {2014})}\BibitemShut {NoStop}%
\bibitem [{\citenamefont {Fourn{\'e}e}\ \emph {et~al.}(2014)\citenamefont
  {Fourn{\'e}e}, \citenamefont {Gaudry}, \citenamefont {Ledieu}, \citenamefont
  {De~Weerd}, \citenamefont {Wu},\ and\ \citenamefont
  {Lograsso}}]{fournee2014self}%
  \BibitemOpen
  \bibfield  {author} {\bibinfo {author} {\bibfnamefont {V.}~\bibnamefont
  {Fourn{\'e}e}}, \bibinfo {author} {\bibfnamefont {{\'E}.}~\bibnamefont
  {Gaudry}}, \bibinfo {author} {\bibfnamefont {J.}~\bibnamefont {Ledieu}},
  \bibinfo {author} {\bibfnamefont {M.-C.}\ \bibnamefont {De~Weerd}}, \bibinfo
  {author} {\bibfnamefont {D.}~\bibnamefont {Wu}}, \ and\ \bibinfo {author}
  {\bibfnamefont {T.}~\bibnamefont {Lograsso}},\ }\href@noop {} {\bibfield
  {journal} {\bibinfo  {journal} {ACS nano}\ }\textbf {\bibinfo {volume} {8}},\
  \bibinfo {pages} {3646} (\bibinfo {year} {2014})}\BibitemShut {NoStop}%
\bibitem [{\citenamefont {Mikhael}\ \emph {et~al.}(2010)\citenamefont
  {Mikhael}, \citenamefont {Schmiedeberg}, \citenamefont {Rausch},
  \citenamefont {Roth}, \citenamefont {Stark},\ and\ \citenamefont
  {Bechinger}}]{mikhael2010proliferation}%
  \BibitemOpen
  \bibfield  {author} {\bibinfo {author} {\bibfnamefont {J.}~\bibnamefont
  {Mikhael}}, \bibinfo {author} {\bibfnamefont {M.}~\bibnamefont
  {Schmiedeberg}}, \bibinfo {author} {\bibfnamefont {S.}~\bibnamefont
  {Rausch}}, \bibinfo {author} {\bibfnamefont {J.}~\bibnamefont {Roth}},
  \bibinfo {author} {\bibfnamefont {H.}~\bibnamefont {Stark}}, \ and\ \bibinfo
  {author} {\bibfnamefont {C.}~\bibnamefont {Bechinger}},\ }\href@noop {}
  {\bibfield  {journal} {\bibinfo  {journal} {Proc. Natl. Acad. Sci. U.S.A.}\
  }\textbf {\bibinfo {volume} {107}},\ \bibinfo {pages} {7214} (\bibinfo {year}
  {2010})}\BibitemShut {NoStop}%
\bibitem [{\citenamefont {Schmiedeberg}\ and\ \citenamefont
  {Stark}(2008)}]{schmiedeberg2008colloidal}%
  \BibitemOpen
  \bibfield  {author} {\bibinfo {author} {\bibfnamefont {M.}~\bibnamefont
  {Schmiedeberg}}\ and\ \bibinfo {author} {\bibfnamefont {H.}~\bibnamefont
  {Stark}},\ }\href@noop {} {\bibfield  {journal} {\bibinfo  {journal} {Phys.
  Rev. Lett.}\ }\textbf {\bibinfo {volume} {101}},\ \bibinfo {pages} {218302}
  (\bibinfo {year} {2008})}\BibitemShut {NoStop}%
\bibitem [{\citenamefont {Schmiedeberg}\ \emph {et~al.}(2010)\citenamefont
  {Schmiedeberg}, \citenamefont {Mikhael}, \citenamefont {Rausch},
  \citenamefont {Roth}, \citenamefont {Helden}, \citenamefont {Bechinger},\
  and\ \citenamefont {Stark}}]{schmiedeberg2010archimedean}%
  \BibitemOpen
  \bibfield  {author} {\bibinfo {author} {\bibfnamefont {M.}~\bibnamefont
  {Schmiedeberg}}, \bibinfo {author} {\bibfnamefont {J.}~\bibnamefont
  {Mikhael}}, \bibinfo {author} {\bibfnamefont {S.}~\bibnamefont {Rausch}},
  \bibinfo {author} {\bibfnamefont {J.}~\bibnamefont {Roth}}, \bibinfo {author}
  {\bibfnamefont {L.}~\bibnamefont {Helden}}, \bibinfo {author} {\bibfnamefont
  {C.}~\bibnamefont {Bechinger}}, \ and\ \bibinfo {author} {\bibfnamefont
  {H.}~\bibnamefont {Stark}},\ }\href@noop {} {\bibfield  {journal} {\bibinfo
  {journal} {Eur. Phys. J. E}\ }\textbf {\bibinfo {volume} {32}},\ \bibinfo
  {pages} {25} (\bibinfo {year} {2010})}\BibitemShut {NoStop}%
\bibitem [{\citenamefont {He}\ \emph {et~al.}(2005)\citenamefont {He},
  \citenamefont {Chen}, \citenamefont {Liu}, \citenamefont {Ribbe},\ and\
  \citenamefont {Mao}}]{he2005self}%
  \BibitemOpen
  \bibfield  {author} {\bibinfo {author} {\bibfnamefont {Y.}~\bibnamefont
  {He}}, \bibinfo {author} {\bibfnamefont {Y.}~\bibnamefont {Chen}}, \bibinfo
  {author} {\bibfnamefont {H.}~\bibnamefont {Liu}}, \bibinfo {author}
  {\bibfnamefont {A.~E.}\ \bibnamefont {Ribbe}}, \ and\ \bibinfo {author}
  {\bibfnamefont {C.}~\bibnamefont {Mao}},\ }\href@noop {} {\bibfield
  {journal} {\bibinfo  {journal} {J. Am. Chem. Soc.}\ }\textbf {\bibinfo
  {volume} {127}},\ \bibinfo {pages} {12202} (\bibinfo {year}
  {2005})}\BibitemShut {NoStop}%
\bibitem [{\citenamefont {Winfree}\ \emph {et~al.}(1998)\citenamefont
  {Winfree}, \citenamefont {Liu}, \citenamefont {Wenzler},\ and\ \citenamefont
  {Seeman}}]{winfree1998design}%
  \BibitemOpen
  \bibfield  {author} {\bibinfo {author} {\bibfnamefont {E.}~\bibnamefont
  {Winfree}}, \bibinfo {author} {\bibfnamefont {F.}~\bibnamefont {Liu}},
  \bibinfo {author} {\bibfnamefont {L.~A.}\ \bibnamefont {Wenzler}}, \ and\
  \bibinfo {author} {\bibfnamefont {N.~C.}\ \bibnamefont {Seeman}},\
  }\href@noop {} {\bibfield  {journal} {\bibinfo  {journal} {Nature}\ }\textbf
  {\bibinfo {volume} {394}},\ \bibinfo {pages} {539} (\bibinfo {year}
  {1998})}\BibitemShut {NoStop}%
\bibitem [{\citenamefont {Bai}\ and\ \citenamefont
  {Abbott}(2010)}]{bai2010recent}%
  \BibitemOpen
  \bibfield  {author} {\bibinfo {author} {\bibfnamefont {Y.}~\bibnamefont
  {Bai}}\ and\ \bibinfo {author} {\bibfnamefont {N.~L.}\ \bibnamefont
  {Abbott}},\ }\href@noop {} {\bibfield  {journal} {\bibinfo  {journal}
  {Langmuir}\ }\textbf {\bibinfo {volume} {27}},\ \bibinfo {pages} {5719}
  (\bibinfo {year} {2010})}\BibitemShut {NoStop}%
\bibitem [{\citenamefont {de~Jeu}, \citenamefont {Ostrovskii},\ and\
  \citenamefont {Shalaginov}(2003)}]{de2003structure}%
  \BibitemOpen
  \bibfield  {author} {\bibinfo {author} {\bibfnamefont {W.~H.}\ \bibnamefont
  {de~Jeu}}, \bibinfo {author} {\bibfnamefont {B.~I.}\ \bibnamefont
  {Ostrovskii}}, \ and\ \bibinfo {author} {\bibfnamefont {A.~N.}\ \bibnamefont
  {Shalaginov}},\ }\href@noop {} {\bibfield  {journal} {\bibinfo  {journal}
  {Rev. Mod. Phys.}\ }\textbf {\bibinfo {volume} {75}},\ \bibinfo {pages} {181}
  (\bibinfo {year} {2003})}\BibitemShut {NoStop}%
\bibitem [{\citenamefont {Mu{\v{s}}evi{\v{c}}}\ \emph
  {et~al.}(2006)\citenamefont {Mu{\v{s}}evi{\v{c}}}, \citenamefont
  {{\v{S}}karabot}, \citenamefont {Tkalec}, \citenamefont {Ravnik},\ and\
  \citenamefont {{\v{Z}}umer}}]{muvsevivc2006two}%
  \BibitemOpen
  \bibfield  {author} {\bibinfo {author} {\bibfnamefont {I.}~\bibnamefont
  {Mu{\v{s}}evi{\v{c}}}}, \bibinfo {author} {\bibfnamefont {M.}~\bibnamefont
  {{\v{S}}karabot}}, \bibinfo {author} {\bibfnamefont {U.}~\bibnamefont
  {Tkalec}}, \bibinfo {author} {\bibfnamefont {M.}~\bibnamefont {Ravnik}}, \
  and\ \bibinfo {author} {\bibfnamefont {S.}~\bibnamefont {{\v{Z}}umer}},\
  }\href@noop {} {\bibfield  {journal} {\bibinfo  {journal} {Science}\ }\textbf
  {\bibinfo {volume} {313}},\ \bibinfo {pages} {954} (\bibinfo {year}
  {2006})}\BibitemShut {NoStop}%
\bibitem [{\citenamefont {Mulero}(2008)}]{mulero2008theory}%
  \BibitemOpen
  \bibfield  {author} {\bibinfo {author} {\bibfnamefont {A.}~\bibnamefont
  {Mulero}},\ }\href@noop {} {\emph {\bibinfo {title} {Theory and simulation of
  hard-sphere fluids and related systems}}},\ Vol.\ \bibinfo {volume} {753}\
  (\bibinfo  {publisher} {Springer Science \& Business Media},\ \bibinfo {year}
  {2008})\BibitemShut {NoStop}%
\bibitem [{\citenamefont {Haji-Akbari}\ \emph {et~al.}(2009)\citenamefont
  {Haji-Akbari}, \citenamefont {Engel}, \citenamefont {Keys}, \citenamefont
  {Zheng}, \citenamefont {Petschek}, \citenamefont {Palffy-Muhoray},\ and\
  \citenamefont {Glotzer}}]{haji2009disordered}%
  \BibitemOpen
  \bibfield  {author} {\bibinfo {author} {\bibfnamefont {A.}~\bibnamefont
  {Haji-Akbari}}, \bibinfo {author} {\bibfnamefont {M.}~\bibnamefont {Engel}},
  \bibinfo {author} {\bibfnamefont {A.~S.}\ \bibnamefont {Keys}}, \bibinfo
  {author} {\bibfnamefont {X.}~\bibnamefont {Zheng}}, \bibinfo {author}
  {\bibfnamefont {R.~G.}\ \bibnamefont {Petschek}}, \bibinfo {author}
  {\bibfnamefont {P.}~\bibnamefont {Palffy-Muhoray}}, \ and\ \bibinfo {author}
  {\bibfnamefont {S.~C.}\ \bibnamefont {Glotzer}},\ }\href@noop {} {\bibfield
  {journal} {\bibinfo  {journal} {Nature}\ }\textbf {\bibinfo {volume} {462}},\
  \bibinfo {pages} {773} (\bibinfo {year} {2009})}\BibitemShut {NoStop}%
\bibitem [{\citenamefont {Bates}\ and\ \citenamefont
  {Frenkel}(2000)}]{Bates2000}%
  \BibitemOpen
  \bibfield  {author} {\bibinfo {author} {\bibfnamefont {M.}~\bibnamefont
  {Bates}}\ and\ \bibinfo {author} {\bibfnamefont {D.}~\bibnamefont
  {Frenkel}},\ }\href {\doibase 10.1063/1.481637} {\bibfield  {journal}
  {\bibinfo  {journal} {J. Chem. Phys.}\ }\textbf {\bibinfo {volume} {112}},\
  \bibinfo {pages} {10034} (\bibinfo {year} {2000})}\BibitemShut {NoStop}%
\bibitem [{\citenamefont {Ghosh}\ and\ \citenamefont
  {Dhar}(2007)}]{ghosh2007orientational}%
  \BibitemOpen
  \bibfield  {author} {\bibinfo {author} {\bibfnamefont {A.}~\bibnamefont
  {Ghosh}}\ and\ \bibinfo {author} {\bibfnamefont {D.}~\bibnamefont {Dhar}},\
  }\href@noop {} {\bibfield  {journal} {\bibinfo  {journal} {EPL}\ }\textbf
  {\bibinfo {volume} {78}},\ \bibinfo {pages} {20003} (\bibinfo {year}
  {2007})}\BibitemShut {NoStop}%
\bibitem [{\citenamefont {K{\"a}hlitz}\ and\ \citenamefont
  {Stark}(2012)}]{kahlitz2012phase}%
  \BibitemOpen
  \bibfield  {author} {\bibinfo {author} {\bibfnamefont {P.}~\bibnamefont
  {K{\"a}hlitz}}\ and\ \bibinfo {author} {\bibfnamefont {H.}~\bibnamefont
  {Stark}},\ }\href@noop {} {\bibfield  {journal} {\bibinfo  {journal} {J.
  Chem. Phys.}\ }\textbf {\bibinfo {volume} {136}},\ \bibinfo {pages} {174705}
  (\bibinfo {year} {2012})}\BibitemShut {NoStop}%
\bibitem [{\citenamefont {Donev}\ \emph {et~al.}(2006)\citenamefont {Donev},
  \citenamefont {Burton}, \citenamefont {Stillinger},\ and\ \citenamefont
  {Torquato}}]{donev2006tetratic}%
  \BibitemOpen
  \bibfield  {author} {\bibinfo {author} {\bibfnamefont {A.}~\bibnamefont
  {Donev}}, \bibinfo {author} {\bibfnamefont {J.}~\bibnamefont {Burton}},
  \bibinfo {author} {\bibfnamefont {F.~H.}\ \bibnamefont {Stillinger}}, \ and\
  \bibinfo {author} {\bibfnamefont {S.}~\bibnamefont {Torquato}},\ }\href@noop
  {} {\bibfield  {journal} {\bibinfo  {journal} {Phys. Rev. B}\ }\textbf
  {\bibinfo {volume} {73}},\ \bibinfo {pages} {054109} (\bibinfo {year}
  {2006})}\BibitemShut {NoStop}%
\bibitem [{\citenamefont {Mart{\'\i}nez-Rat{\'o}n}, \citenamefont {Velasco},\
  and\ \citenamefont {Mederos}(2005)}]{martinez2005effect}%
  \BibitemOpen
  \bibfield  {author} {\bibinfo {author} {\bibfnamefont {Y.}~\bibnamefont
  {Mart{\'\i}nez-Rat{\'o}n}}, \bibinfo {author} {\bibfnamefont
  {E.}~\bibnamefont {Velasco}}, \ and\ \bibinfo {author} {\bibfnamefont
  {L.}~\bibnamefont {Mederos}},\ }\href@noop {} {\bibfield  {journal} {\bibinfo
   {journal} {J. Chem. Phys.}\ }\textbf {\bibinfo {volume} {122}},\ \bibinfo
  {pages} {064903} (\bibinfo {year} {2005})}\BibitemShut {NoStop}%
\bibitem [{\citenamefont {Lagomarsino}, \citenamefont {Dogterom},\ and\
  \citenamefont {Dijkstra}(2003)}]{lagomarsino2003isotropic}%
  \BibitemOpen
  \bibfield  {author} {\bibinfo {author} {\bibfnamefont {M.~C.}\ \bibnamefont
  {Lagomarsino}}, \bibinfo {author} {\bibfnamefont {M.}~\bibnamefont
  {Dogterom}}, \ and\ \bibinfo {author} {\bibfnamefont {M.}~\bibnamefont
  {Dijkstra}},\ }\href@noop {} {\bibfield  {journal} {\bibinfo  {journal} {J.
  Chem. Phys.}\ }\textbf {\bibinfo {volume} {119}},\ \bibinfo {pages} {3535}
  (\bibinfo {year} {2003})}\BibitemShut {NoStop}%
\bibitem [{\citenamefont {K{\"a}hlitz}, \citenamefont {Schoen},\ and\
  \citenamefont {Stark}(2012)}]{kahlitz2012clustering}%
  \BibitemOpen
  \bibfield  {author} {\bibinfo {author} {\bibfnamefont {P.}~\bibnamefont
  {K{\"a}hlitz}}, \bibinfo {author} {\bibfnamefont {M.}~\bibnamefont {Schoen}},
  \ and\ \bibinfo {author} {\bibfnamefont {H.}~\bibnamefont {Stark}},\
  }\href@noop {} {\bibfield  {journal} {\bibinfo  {journal} {J. Chem. Phys.}\
  }\textbf {\bibinfo {volume} {137}},\ \bibinfo {pages} {224705} (\bibinfo
  {year} {2012})}\BibitemShut {NoStop}%
\bibitem [{\citenamefont {Xu}\ \emph {et~al.}(2013)\citenamefont {Xu},
  \citenamefont {Li}, \citenamefont {Sun},\ and\ \citenamefont
  {An}}]{xu2013hard}%
  \BibitemOpen
  \bibfield  {author} {\bibinfo {author} {\bibfnamefont {W.-S.}\ \bibnamefont
  {Xu}}, \bibinfo {author} {\bibfnamefont {Y.-W.}\ \bibnamefont {Li}}, \bibinfo
  {author} {\bibfnamefont {Z.-Y.}\ \bibnamefont {Sun}}, \ and\ \bibinfo
  {author} {\bibfnamefont {L.-J.}\ \bibnamefont {An}},\ }\href@noop {}
  {\bibfield  {journal} {\bibinfo  {journal} {J. Chem. Phys.}\ }\textbf
  {\bibinfo {volume} {139}},\ \bibinfo {pages} {024501} (\bibinfo {year}
  {2013})}\BibitemShut {NoStop}%
\bibitem [{\citenamefont {Moradi}, \citenamefont {Hashemi},\ and\ \citenamefont
  {Taghizadeh}(2010)}]{moradi2010monte}%
  \BibitemOpen
  \bibfield  {author} {\bibinfo {author} {\bibfnamefont {M.}~\bibnamefont
  {Moradi}}, \bibinfo {author} {\bibfnamefont {S.}~\bibnamefont {Hashemi}}, \
  and\ \bibinfo {author} {\bibfnamefont {F.}~\bibnamefont {Taghizadeh}},\
  }\href@noop {} {\bibfield  {journal} {\bibinfo  {journal} {Physica A}\
  }\textbf {\bibinfo {volume} {389}},\ \bibinfo {pages} {4510} (\bibinfo {year}
  {2010})}\BibitemShut {NoStop}%
\bibitem [{\citenamefont {Care}\ and\ \citenamefont
  {Cleaver}(2005)}]{care2005computer}%
  \BibitemOpen
  \bibfield  {author} {\bibinfo {author} {\bibfnamefont {C.}~\bibnamefont
  {Care}}\ and\ \bibinfo {author} {\bibfnamefont {D.}~\bibnamefont {Cleaver}},\
  }\href@noop {} {\bibfield  {journal} {\bibinfo  {journal} {Rep. Prog. Phys.}\
  }\textbf {\bibinfo {volume} {68}},\ \bibinfo {pages} {2665} (\bibinfo {year}
  {2005})}\BibitemShut {NoStop}%
\bibitem [{\citenamefont {Lubensky}(2002)}]{Lubensky2002}%
  \BibitemOpen
  \bibfield  {author} {\bibinfo {author} {\bibfnamefont {T.~C.}\ \bibnamefont
  {Lubensky}},\ }\href {\doibase 10.1103/PhysRevE.66.031704} {\bibfield
  {journal} {\bibinfo  {journal} {Phys. Rev. E}\ }\textbf {\bibinfo {volume}
  {66}},\ \bibinfo {pages} {031704} (\bibinfo {year} {2002})}\BibitemShut
  {NoStop}%
\bibitem [{\citenamefont {Takezoe}\ and\ \citenamefont
  {Takanishi}(2006)}]{Takezoe2006}%
  \BibitemOpen
  \bibfield  {author} {\bibinfo {author} {\bibfnamefont {H.}~\bibnamefont
  {Takezoe}}\ and\ \bibinfo {author} {\bibfnamefont {Y.}~\bibnamefont
  {Takanishi}},\ }\href {\doibase 10.1143/JJAP.45.597} {\bibfield  {journal}
  {\bibinfo  {journal} {Jpn. J. Appl. Phys.}\ }\textbf {\bibinfo {volume}
  {45}},\ \bibinfo {pages} {597} (\bibinfo {year} {2006})}\BibitemShut
  {NoStop}%
\bibitem [{\citenamefont {Pelzl}, \citenamefont {Diele},\ and\ \citenamefont
  {Weissflog}(1999)}]{Pelzl1999}%
  \BibitemOpen
  \bibfield  {author} {\bibinfo {author} {\bibfnamefont {G.}~\bibnamefont
  {Pelzl}}, \bibinfo {author} {\bibfnamefont {S.}~\bibnamefont {Diele}}, \ and\
  \bibinfo {author} {\bibfnamefont {W.}~\bibnamefont {Weissflog}},\ }\href
  {http://onlinelibrary.wiley.com/doi/10.1002/(SICI)1521-4095(199906)11:9\%3C707::AID-ADMA707\%3E3.0.CO;2-D/abstract}
  {\bibfield  {journal} {\bibinfo  {journal} {Adv. Mater.}\ ,\ \bibinfo {pages}
  {707}} (\bibinfo {year} {1999})}\BibitemShut {NoStop}%
\bibitem [{\citenamefont {Ros}\ \emph {et~al.}(2005)\citenamefont {Ros},
  \citenamefont {Serrano}, \citenamefont {de~la Fuente},\ and\ \citenamefont
  {Folcia}}]{Ros2005}%
  \BibitemOpen
  \bibfield  {author} {\bibinfo {author} {\bibfnamefont {M.~B.}\ \bibnamefont
  {Ros}}, \bibinfo {author} {\bibfnamefont {J.~L.}\ \bibnamefont {Serrano}},
  \bibinfo {author} {\bibfnamefont {M.~R.}\ \bibnamefont {de~la Fuente}}, \
  and\ \bibinfo {author} {\bibfnamefont {C.~L.}\ \bibnamefont {Folcia}},\
  }\href {\doibase 10.1039/b504384k} {\bibfield  {journal} {\bibinfo  {journal}
  {J. Mater. Chem.}\ }\textbf {\bibinfo {volume} {15}},\ \bibinfo {pages}
  {5093} (\bibinfo {year} {2005})}\BibitemShut {NoStop}%
\bibitem [{\citenamefont {Berardi}\ \emph {et~al.}(2008)\citenamefont
  {Berardi}, \citenamefont {Muccioli}, \citenamefont {Orlandi}, \citenamefont
  {Ricci},\ and\ \citenamefont {Zannoni}}]{Berardi2008}%
  \BibitemOpen
  \bibfield  {author} {\bibinfo {author} {\bibfnamefont {R.}~\bibnamefont
  {Berardi}}, \bibinfo {author} {\bibfnamefont {L.}~\bibnamefont {Muccioli}},
  \bibinfo {author} {\bibfnamefont {S.}~\bibnamefont {Orlandi}}, \bibinfo
  {author} {\bibfnamefont {M.}~\bibnamefont {Ricci}}, \ and\ \bibinfo {author}
  {\bibfnamefont {C.}~\bibnamefont {Zannoni}},\ }\href {\doibase
  10.1088/0953-8984/20/46/463101} {\bibfield  {journal} {\bibinfo  {journal}
  {J. Phys. Cond. Matt.}\ }\textbf {\bibinfo {volume} {20}},\ \bibinfo {pages}
  {463101} (\bibinfo {year} {2008})}\BibitemShut {NoStop}%
\bibitem [{\citenamefont {Teixeira}, \citenamefont {Masters},\ and\
  \citenamefont {Mulder}(1998)}]{Teixeira1998}%
  \BibitemOpen
  \bibfield  {author} {\bibinfo {author} {\bibfnamefont {P.~I.~C.}\
  \bibnamefont {Teixeira}}, \bibinfo {author} {\bibfnamefont {J.}~\bibnamefont
  {Masters}}, \ and\ \bibinfo {author} {\bibfnamefont {B.~M.}\ \bibnamefont
  {Mulder}},\ }\href {\doibase 10.1080/10587259808048440} {\bibfield  {journal}
  {\bibinfo  {journal} {Mol. Cryst. Liq. Cryst.}\ }\textbf {\bibinfo {volume}
  {323}},\ \bibinfo {pages} {167} (\bibinfo {year} {1998})}\BibitemShut
  {NoStop}%
\bibitem [{\citenamefont {Pelzl}\ \emph {et~al.}(2002)\citenamefont {Pelzl},
  \citenamefont {Eremin}, \citenamefont {Diele}, \citenamefont {Kresse},\ and\
  \citenamefont {Weissflog}}]{Pelzl2002}%
  \BibitemOpen
  \bibfield  {author} {\bibinfo {author} {\bibfnamefont {G.}~\bibnamefont
  {Pelzl}}, \bibinfo {author} {\bibfnamefont {A.}~\bibnamefont {Eremin}},
  \bibinfo {author} {\bibfnamefont {S.}~\bibnamefont {Diele}}, \bibinfo
  {author} {\bibfnamefont {H.}~\bibnamefont {Kresse}}, \ and\ \bibinfo {author}
  {\bibfnamefont {W.}~\bibnamefont {Weissflog}},\ }\href {\doibase
  10.1039/b206236d} {\bibfield  {journal} {\bibinfo  {journal} {J. Mater.
  Chem.}\ }\textbf {\bibinfo {volume} {12}},\ \bibinfo {pages} {2591} (\bibinfo
  {year} {2002})}\BibitemShut {NoStop}%
\bibitem [{\citenamefont {Dozov}(2001)}]{Dozov2001a}%
  \BibitemOpen
  \bibfield  {author} {\bibinfo {author} {\bibfnamefont {I.}~\bibnamefont
  {Dozov}},\ }\href {http://iopscience.iop.org/0295-5075/56/2/247} {\bibfield
  {journal} {\bibinfo  {journal} {EPL}\ }\textbf {\bibinfo {volume} {247}}
  (\bibinfo {year} {2001})}\BibitemShut {NoStop}%
\bibitem [{\citenamefont {Camp}, \citenamefont {Allen},\ and\ \citenamefont
  {Masters}(1999)}]{Camp1999a}%
  \BibitemOpen
  \bibfield  {author} {\bibinfo {author} {\bibfnamefont {P.~J.}\ \bibnamefont
  {Camp}}, \bibinfo {author} {\bibfnamefont {M.~P.}\ \bibnamefont {Allen}}, \
  and\ \bibinfo {author} {\bibfnamefont {A.~J.}\ \bibnamefont {Masters}},\
  }\href {\doibase 10.1063/1.480324} {\bibfield  {journal} {\bibinfo  {journal}
  {J. Chem. Phys.}\ }\textbf {\bibinfo {volume} {111}},\ \bibinfo {pages}
  {9871} (\bibinfo {year} {1999})}\BibitemShut {NoStop}%
\bibitem [{\citenamefont {Johnston}, \citenamefont {Low},\ and\ \citenamefont
  {Neal}(2002)}]{johnston2002computer}%
  \BibitemOpen
  \bibfield  {author} {\bibinfo {author} {\bibfnamefont {S.~J.}\ \bibnamefont
  {Johnston}}, \bibinfo {author} {\bibfnamefont {R.~J.}\ \bibnamefont {Low}}, \
  and\ \bibinfo {author} {\bibfnamefont {M.~P.}\ \bibnamefont {Neal}},\
  }\href@noop {} {\bibfield  {journal} {\bibinfo  {journal} {Phys. Rev. E}\
  }\textbf {\bibinfo {volume} {65}},\ \bibinfo {pages} {051706} (\bibinfo
  {year} {2002})}\BibitemShut {NoStop}%
\bibitem [{\citenamefont {Dewar}\ and\ \citenamefont
  {Camp}(2004)}]{dewar2004computer}%
  \BibitemOpen
  \bibfield  {author} {\bibinfo {author} {\bibfnamefont {A.}~\bibnamefont
  {Dewar}}\ and\ \bibinfo {author} {\bibfnamefont {P.~J.}\ \bibnamefont
  {Camp}},\ }\href@noop {} {\bibfield  {journal} {\bibinfo  {journal} {Phys.
  Rev. E}\ }\textbf {\bibinfo {volume} {70}},\ \bibinfo {pages} {011704}
  (\bibinfo {year} {2004})}\BibitemShut {NoStop}%
\bibitem [{\citenamefont {Shamid}, \citenamefont {Dhakal},\ and\ \citenamefont
  {Selinger}(2013)}]{shamid2013statistical}%
  \BibitemOpen
  \bibfield  {author} {\bibinfo {author} {\bibfnamefont {S.~M.}\ \bibnamefont
  {Shamid}}, \bibinfo {author} {\bibfnamefont {S.}~\bibnamefont {Dhakal}}, \
  and\ \bibinfo {author} {\bibfnamefont {J.~V.}\ \bibnamefont {Selinger}},\
  }\href@noop {} {\bibfield  {journal} {\bibinfo  {journal} {Phys. Rev. E}\
  }\textbf {\bibinfo {volume} {87}},\ \bibinfo {pages} {052503} (\bibinfo
  {year} {2013})}\BibitemShut {NoStop}%
\bibitem [{\citenamefont {Grzybowski}\ and\ \citenamefont
  {Longa}(2011)}]{grzybowski2011biaxial}%
  \BibitemOpen
  \bibfield  {author} {\bibinfo {author} {\bibfnamefont {P.}~\bibnamefont
  {Grzybowski}}\ and\ \bibinfo {author} {\bibfnamefont {L.}~\bibnamefont
  {Longa}},\ }\href@noop {} {\bibfield  {journal} {\bibinfo  {journal} {Phys.
  Rev. Lett.}\ }\textbf {\bibinfo {volume} {107}},\ \bibinfo {pages} {027802}
  (\bibinfo {year} {2011})}\BibitemShut {NoStop}%
\bibitem [{\citenamefont {Mart\'{\i}nez-Gonz\'{a}lez}, \citenamefont
  {Armas-P\'{e}rez},\ and\ \citenamefont
  {Quintana-H}(2012)}]{Martinez-Gonzalez2012}%
  \BibitemOpen
  \bibfield  {author} {\bibinfo {author} {\bibfnamefont {J.}~\bibnamefont
  {Mart\'{\i}nez-Gonz\'{a}lez}}, \bibinfo {author} {\bibfnamefont {J.~C.}\
  \bibnamefont {Armas-P\'{e}rez}}, \ and\ \bibinfo {author} {\bibfnamefont
  {J.}~\bibnamefont {Quintana-H}},\ }\href {\doibase 10.1007/s10955-012-0606-7}
  {\bibfield  {journal} {\bibinfo  {journal} {J. Stat. Phys.}\ }\textbf
  {\bibinfo {volume} {150}},\ \bibinfo {pages} {559} (\bibinfo {year}
  {2012})}\BibitemShut {NoStop}%
\bibitem [{\citenamefont {Bisi}\ \emph {et~al.}(2008)\citenamefont {Bisi},
  \citenamefont {Rosso}, \citenamefont {Virga},\ and\ \citenamefont
  {Durand}}]{Bisi2008}%
  \BibitemOpen
  \bibfield  {author} {\bibinfo {author} {\bibfnamefont {F.}~\bibnamefont
  {Bisi}}, \bibinfo {author} {\bibfnamefont {R.}~\bibnamefont {Rosso}},
  \bibinfo {author} {\bibfnamefont {E.}~\bibnamefont {Virga}}, \ and\ \bibinfo
  {author} {\bibfnamefont {G.}~\bibnamefont {Durand}},\ }\href {\doibase
  10.1103/PhysRevE.78.011705} {\bibfield  {journal} {\bibinfo  {journal} {Phys.
  Rev. E}\ }\textbf {\bibinfo {volume} {78}},\ \bibinfo {pages} {011705}
  (\bibinfo {year} {2008})}\BibitemShut {NoStop}%
\bibitem [{\citenamefont {Mart\'{\i}nez-Gonz\'{a}lez}\ \emph
  {et~al.}(2012)\citenamefont {Mart\'{\i}nez-Gonz\'{a}lez}, \citenamefont
  {Varga}, \citenamefont {Gurin},\ and\ \citenamefont
  {Quintana-H.}}]{Martinez-Gonzalez2012a}%
  \BibitemOpen
  \bibfield  {author} {\bibinfo {author} {\bibfnamefont {J.}~\bibnamefont
  {Mart\'{\i}nez-Gonz\'{a}lez}}, \bibinfo {author} {\bibfnamefont
  {S.}~\bibnamefont {Varga}}, \bibinfo {author} {\bibfnamefont
  {P.}~\bibnamefont {Gurin}}, \ and\ \bibinfo {author} {\bibfnamefont
  {J.}~\bibnamefont {Quintana-H.}},\ }\href {\doibase
  10.1209/0295-5075/97/26004} {\bibfield  {journal} {\bibinfo  {journal} {EPL}\
  }\textbf {\bibinfo {volume} {97}},\ \bibinfo {pages} {26004} (\bibinfo {year}
  {2012})}\BibitemShut {NoStop}%
\bibitem [{\citenamefont {Pe{\'o}n}\ \emph {et~al.}(2006)\citenamefont
  {Pe{\'o}n}, \citenamefont {Saucedo-Zugazagoitia}, \citenamefont
  {Pucheta-Mendez}, \citenamefont {Perusqu{\'\i}a}, \citenamefont {Sutmann},\
  and\ \citenamefont {Quintana-H}}]{Peon2006}%
  \BibitemOpen
  \bibfield  {author} {\bibinfo {author} {\bibfnamefont {J.}~\bibnamefont
  {Pe{\'o}n}}, \bibinfo {author} {\bibfnamefont {J.}~\bibnamefont
  {Saucedo-Zugazagoitia}}, \bibinfo {author} {\bibfnamefont {F.}~\bibnamefont
  {Pucheta-Mendez}}, \bibinfo {author} {\bibfnamefont {R.~A.}\ \bibnamefont
  {Perusqu{\'\i}a}}, \bibinfo {author} {\bibfnamefont {G.}~\bibnamefont
  {Sutmann}}, \ and\ \bibinfo {author} {\bibfnamefont {J.}~\bibnamefont
  {Quintana-H}},\ }\href {\doibase 10.1063/1.2338313} {\bibfield  {journal}
  {\bibinfo  {journal} {J. Chem. Phys.}\ }\textbf {\bibinfo {volume} {125}},\
  \bibinfo {pages} {104908} (\bibinfo {year} {2006})}\BibitemShut {NoStop}%
\bibitem [{\citenamefont {Mermin}(1968)}]{PhysRev.176.250}%
  \BibitemOpen
  \bibfield  {author} {\bibinfo {author} {\bibfnamefont {N.~D.}\ \bibnamefont
  {Mermin}},\ }\href {\doibase 10.1103/PhysRev.176.250} {\bibfield  {journal}
  {\bibinfo  {journal} {Phys. Rev.}\ }\textbf {\bibinfo {volume} {176}},\
  \bibinfo {pages} {250} (\bibinfo {year} {1968})}\BibitemShut {NoStop}%
\bibitem [{\citenamefont {Strandburg}(1988)}]{strandburg1988two}%
  \BibitemOpen
  \bibfield  {author} {\bibinfo {author} {\bibfnamefont {K.~J.}\ \bibnamefont
  {Strandburg}},\ }\href@noop {} {\bibfield  {journal} {\bibinfo  {journal}
  {Rev. Mod. Phys.}\ }\textbf {\bibinfo {volume} {60}},\ \bibinfo {pages} {161}
  (\bibinfo {year} {1988})}\BibitemShut {NoStop}%
\bibitem [{\citenamefont {Bernard}\ and\ \citenamefont
  {Krauth}(2011)}]{bernard2011two}%
  \BibitemOpen
  \bibfield  {author} {\bibinfo {author} {\bibfnamefont {E.~P.}\ \bibnamefont
  {Bernard}}\ and\ \bibinfo {author} {\bibfnamefont {W.}~\bibnamefont
  {Krauth}},\ }\href@noop {} {\bibfield  {journal} {\bibinfo  {journal} {Phys.
  Rev. Lett.}\ }\textbf {\bibinfo {volume} {107}},\ \bibinfo {pages} {155704}
  (\bibinfo {year} {2011})}\BibitemShut {NoStop}%
\bibitem [{\citenamefont {Borshch}\ \emph {et~al.}(2013)\citenamefont
  {Borshch}, \citenamefont {Kim}, \citenamefont {Xiang}, \citenamefont {Gao},
  \citenamefont {J{\'a}kli}, \citenamefont {Panov}, \citenamefont {Vij},
  \citenamefont {Imrie}, \citenamefont {Tamba}, \citenamefont {Mehl} \emph
  {et~al.}}]{borshch2013nematic}%
  \BibitemOpen
  \bibfield  {author} {\bibinfo {author} {\bibfnamefont {V.}~\bibnamefont
  {Borshch}}, \bibinfo {author} {\bibfnamefont {Y.-K.}\ \bibnamefont {Kim}},
  \bibinfo {author} {\bibfnamefont {J.}~\bibnamefont {Xiang}}, \bibinfo
  {author} {\bibfnamefont {M.}~\bibnamefont {Gao}}, \bibinfo {author}
  {\bibfnamefont {A.}~\bibnamefont {J{\'a}kli}}, \bibinfo {author}
  {\bibfnamefont {V.~P.}\ \bibnamefont {Panov}}, \bibinfo {author}
  {\bibfnamefont {J.~K.}\ \bibnamefont {Vij}}, \bibinfo {author} {\bibfnamefont
  {C.~T.}\ \bibnamefont {Imrie}}, \bibinfo {author} {\bibfnamefont {M.-G.}\
  \bibnamefont {Tamba}}, \bibinfo {author} {\bibfnamefont {G.~H.}\ \bibnamefont
  {Mehl}},  \emph {et~al.},\ }\href@noop {} {\bibfield  {journal} {\bibinfo
  {journal} {Nat. commun.}\ }\textbf {\bibinfo {volume} {4}} (\bibinfo {year}
  {2013})}\BibitemShut {NoStop}%
\bibitem [{\citenamefont {Memmer}(2002)}]{memmer2002liquid}%
  \BibitemOpen
  \bibfield  {author} {\bibinfo {author} {\bibfnamefont {R.}~\bibnamefont
  {Memmer}},\ }\href@noop {} {\bibfield  {journal} {\bibinfo  {journal} {Liq.
  Cryst.}\ }\textbf {\bibinfo {volume} {29}},\ \bibinfo {pages} {483} (\bibinfo
  {year} {2002})}\BibitemShut {NoStop}%
\bibitem [{\citenamefont {Frenkel}\ and\ \citenamefont
  {Eppenga}(1985)}]{Frenkel1985}%
  \BibitemOpen
  \bibfield  {author} {\bibinfo {author} {\bibfnamefont {D.}~\bibnamefont
  {Frenkel}}\ and\ \bibinfo {author} {\bibfnamefont {R.}~\bibnamefont
  {Eppenga}},\ }\href {http://pra.aps.org/abstract/PRA/v31/i3/p1776\_1}
  {\bibfield  {journal} {\bibinfo  {journal} {Phys. Rev. A}\ } (\bibinfo {year}
  {1985})}\BibitemShut {NoStop}%
\bibitem [{\citenamefont {Vink}(2009)}]{Vink2009}%
  \BibitemOpen
  \bibfield  {author} {\bibinfo {author} {\bibfnamefont {R.~L.}\ \bibnamefont
  {Vink}},\ }\href {\doibase 10.1140/epjb/e2009-00333-x} {\bibfield  {journal}
  {\bibinfo  {journal} {Eur. Phys. J. B}\ }\textbf {\bibinfo {volume} {72}},\
  \bibinfo {pages} {225} (\bibinfo {year} {2009})}\BibitemShut {NoStop}%
\bibitem [{\citenamefont {Armas-P\'{e}rez}\ and\ \citenamefont
  {Quintana-H.}(2011)}]{Armas-Perez2011}%
  \BibitemOpen
  \bibfield  {author} {\bibinfo {author} {\bibfnamefont {J.~C.}\ \bibnamefont
  {Armas-P\'{e}rez}}\ and\ \bibinfo {author} {\bibfnamefont {J.}~\bibnamefont
  {Quintana-H.}},\ }\href {\doibase 10.1103/PhysRevE.83.051709} {\bibfield
  {journal} {\bibinfo  {journal} {Phys. Rev. E}\ }\textbf {\bibinfo {volume}
  {83}},\ \bibinfo {pages} {051709} (\bibinfo {year} {2011})}\BibitemShut
  {NoStop}%
\bibitem [{\citenamefont {Varga}\ \emph {et~al.}(2009)\citenamefont {Varga},
  \citenamefont {Gurin}, \citenamefont {Armas-P\'{e}rez},\ and\ \citenamefont
  {Quintana-H}}]{Varga2009b}%
  \BibitemOpen
  \bibfield  {author} {\bibinfo {author} {\bibfnamefont {S.}~\bibnamefont
  {Varga}}, \bibinfo {author} {\bibfnamefont {P.}~\bibnamefont {Gurin}},
  \bibinfo {author} {\bibfnamefont {J.~C.}\ \bibnamefont {Armas-P\'{e}rez}}, \
  and\ \bibinfo {author} {\bibfnamefont {J.}~\bibnamefont {Quintana-H}},\
  }\href {\doibase 10.1063/1.3258858} {\bibfield  {journal} {\bibinfo
  {journal} {J. Chem. Phys.}\ }\textbf {\bibinfo {volume} {131}},\ \bibinfo
  {pages} {184901} (\bibinfo {year} {2009})}\BibitemShut {NoStop}%
\bibitem [{\citenamefont {Perusqu{\i}́a}, \citenamefont {Pe{\'o}n},\ and\
  \citenamefont {Quintana}(2005)}]{Perusqua2005}%
  \BibitemOpen
  \bibfield  {author} {\bibinfo {author} {\bibfnamefont {R.~A.}\ \bibnamefont
  {Perusqu{\i}́a}}, \bibinfo {author} {\bibfnamefont {J.}~\bibnamefont
  {Pe{\'o}n}}, \ and\ \bibinfo {author} {\bibfnamefont {J.}~\bibnamefont
  {Quintana}},\ }\href {\doibase 10.1016/j.physa.2004.05.089} {\bibfield
  {journal} {\bibinfo  {journal} {Physica A}\ }\textbf {\bibinfo {volume}
  {345}},\ \bibinfo {pages} {130} (\bibinfo {year} {2005})}\BibitemShut
  {NoStop}%
\bibitem [{\citenamefont {Polson}\ and\ \citenamefont
  {Frenkel}(1997)}]{Polson1997}%
  \BibitemOpen
  \bibfield  {author} {\bibinfo {author} {\bibfnamefont {J.}~\bibnamefont
  {Polson}}\ and\ \bibinfo {author} {\bibfnamefont {D.}~\bibnamefont
  {Frenkel}},\ }\href {\doibase 10.1103/PhysRevE.56.R6260} {\bibfield
  {journal} {\bibinfo  {journal} {Phys. Rev. E}\ }\textbf {\bibinfo {volume}
  {56}},\ \bibinfo {pages} {R6260} (\bibinfo {year} {1997})}\BibitemShut
  {NoStop}%
\bibitem [{\citenamefont {Frenkel}\ and\ \citenamefont
  {Smit}(1997)}]{Frenkel1997}%
  \BibitemOpen
  \bibfield  {author} {\bibinfo {author} {\bibfnamefont {D.}~\bibnamefont
  {Frenkel}}\ and\ \bibinfo {author} {\bibfnamefont {B.}~\bibnamefont {Smit}},\
  } {\enquote {\bibinfo {title}
  {{Understanding Molecular Simulation: From Algorithms to Applications}},}\ }
  (\bibinfo {year} {1997})\BibitemShut {NoStop}%
\bibitem [{\citenamefont {Miller}, \citenamefont {Amon},\ and\ \citenamefont
  {Reinhardt}(2000)}]{Miller2000}%
  \BibitemOpen
  \bibfield  {author} {\bibinfo {author} {\bibfnamefont {M.}~\bibnamefont
  {Miller}}, \bibinfo {author} {\bibfnamefont {L.}~\bibnamefont {Amon}}, \ and\
  \bibinfo {author} {\bibfnamefont {W.}~\bibnamefont {Reinhardt}},\ }\href
  {http://www.sciencedirect.com/science/article/pii/S0009261400012173}
  {\bibfield  {journal} {\bibinfo  {journal} {Chem. Phys. Lett.}\ }\textbf
  {\bibinfo {volume} {331}},\ \bibinfo {pages} {278} (\bibinfo {year}
  {2000})}\BibitemShut {NoStop}%
\bibitem [{\citenamefont {Sutmann}\ and\ \citenamefont
  {Stegailov}(2006)}]{Sutmann2006}%
  \BibitemOpen
  \bibfield  {author} {\bibinfo {author} {\bibfnamefont {G.}~\bibnamefont
  {Sutmann}}\ and\ \bibinfo {author} {\bibfnamefont {V.}~\bibnamefont
  {Stegailov}},\ }\href {\doibase 10.1016/j.molliq.2005.11.029} {\bibfield
  {journal} {\bibinfo  {journal} {J. Mol. Liq.}\ }\textbf {\bibinfo {volume}
  {125}},\ \bibinfo {pages} {197} (\bibinfo {year} {2006})}\BibitemShut
  {NoStop}%
\bibitem [{\citenamefont {Wang}\ and\ \citenamefont
  {Swendsen}(1990)}]{Wang1990}%
  \BibitemOpen
  \bibfield  {author} {\bibinfo {author} {\bibfnamefont {J.}~\bibnamefont
  {Wang}}\ and\ \bibinfo {author} {\bibfnamefont {R.}~\bibnamefont
  {Swendsen}},\ }\href@noop {} {\bibfield  {journal} {\bibinfo  {journal}
  {Physica A}\ }\textbf {\bibinfo {volume} {167}},\ \bibinfo {pages} {565}
  (\bibinfo {year} {1990})}\BibitemShut {NoStop}%
\bibitem [{\citenamefont {Dressts}\ and\ \citenamefont
  {Krauth}(1995)}]{Dressts1995}%
  \BibitemOpen
  \bibfield  {author} {\bibinfo {author} {\bibfnamefont {C.}~\bibnamefont
  {Dressts}}\ and\ \bibinfo {author} {\bibfnamefont {W.}~\bibnamefont
  {Krauth}},\ }\href@noop {} {\bibfield  {journal} {\bibinfo  {journal} {J.
  Phys. A-Math. Gen.}\ }\textbf {\bibinfo {volume} {28}},\ \bibinfo {pages}
  {L597} (\bibinfo {year} {1995})}\BibitemShut {NoStop}%
\bibitem [{\citenamefont {Whitelam}\ and\ \citenamefont
  {Geissler}(2007)}]{Whitelam2007}%
  \BibitemOpen
  \bibfield  {author} {\bibinfo {author} {\bibfnamefont {S.}~\bibnamefont
  {Whitelam}}\ and\ \bibinfo {author} {\bibfnamefont {P.~L.}\ \bibnamefont
  {Geissler}},\ }\href {\doibase 10.1063/1.2790421} {\bibfield  {journal}
  {\bibinfo  {journal} {J. Chem. Phys.}\ }\textbf {\bibinfo {volume} {127}},\
  \bibinfo {pages} {154101} (\bibinfo {year} {2007})}\BibitemShut {NoStop}%
\bibitem [{\citenamefont {Wu}, \citenamefont {Chandler},\ and\ \citenamefont
  {Smitt}(1992)}]{Wu1992}%
  \BibitemOpen
  \bibfield  {author} {\bibinfo {author} {\bibfnamefont {D.}~\bibnamefont
  {Wu}}, \bibinfo {author} {\bibfnamefont {D.}~\bibnamefont {Chandler}}, \ and\
  \bibinfo {author} {\bibfnamefont {B.}~\bibnamefont {Smitt}},\ }\href@noop {}
  {\bibfield  {journal} {\bibinfo  {journal} {J. Phys. Chem.}\ }\textbf
  {\bibinfo {volume} {96}},\ \bibinfo {pages} {4077} (\bibinfo {year}
  {1992})}\BibitemShut {NoStop}%
\bibitem [{\citenamefont {Onsager}(1949)}]{Onsager1949}%
  \BibitemOpen
  \bibfield  {author} {\bibinfo {author} {\bibfnamefont {L.}~\bibnamefont
  {Onsager}},\ }\href
  {http://onlinelibrary.wiley.com/doi/10.1111/j.1749-6632.1949.tb27296.x/full}
  {\bibfield  {journal} {\bibinfo  {journal} {Ann. N. Y. Acad. Sci.}\ }
  (\bibinfo {year} {1949})}\BibitemShut {NoStop}%
\bibitem [{\citenamefont {Mart\'{\i}nez-Rat\'{o}n}, \citenamefont {Velasco},\
  and\ \citenamefont {Mederos}(2005)}]{Martinez-Raton2005}%
  \BibitemOpen
  \bibfield  {author} {\bibinfo {author} {\bibfnamefont {Y.}~\bibnamefont
  {Mart\'{\i}nez-Rat\'{o}n}}, \bibinfo {author} {\bibfnamefont
  {E.}~\bibnamefont {Velasco}}, \ and\ \bibinfo {author} {\bibfnamefont
  {L.}~\bibnamefont {Mederos}},\ }\href {\doibase 10.1063/1.1849159} {\bibfield
   {journal} {\bibinfo  {journal} {J. Chem. Phys.}\ }\textbf {\bibinfo {volume}
  {122}},\ \bibinfo {pages} {064903} (\bibinfo {year} {2005})}\BibitemShut
  {NoStop}%
\bibitem [{\citenamefont {Mermin}\ and\ \citenamefont
  {Wagner}(1966)}]{Merminf1966}%
  \BibitemOpen
  \bibfield  {author} {\bibinfo {author} {\bibfnamefont {N.}~\bibnamefont
  {Mermin}}\ and\ \bibinfo {author} {\bibfnamefont {H.}~\bibnamefont
  {Wagner}},\ }\href {http://adsabs.harvard.edu/abs/1966PhRvL..17.1133M}
  {\bibfield  {journal} {\bibinfo  {journal} {Phys. Rev. Lett.}\ }\textbf
  {\bibinfo {volume} {17}},\ \bibinfo {pages} {1133} (\bibinfo {year}
  {1966})}\BibitemShut {NoStop}%
\bibitem [{\citenamefont {Kosterlitz}\ and\ \citenamefont
  {Thouless}(1973)}]{Kosterlitz1973}%
  \BibitemOpen
  \bibfield  {author} {\bibinfo {author} {\bibfnamefont {J.}~\bibnamefont
  {Kosterlitz}}\ and\ \bibinfo {author} {\bibfnamefont {D.}~\bibnamefont
  {Thouless}},\ }\href {http://iopscience.iop.org/0022-3719/6/7/010} {\bibfield
   {journal} {\bibinfo  {journal} {J. Phys. C Solid State}\ }\textbf {\bibinfo
  {volume} {1181}} (\bibinfo {year} {1973})}\BibitemShut {NoStop}%
\bibitem [{\citenamefont {Nelson}\ and\ \citenamefont
  {Pelcovits}(1977)}]{Nelson1977}%
  \BibitemOpen
  \bibfield  {author} {\bibinfo {author} {\bibfnamefont {D.~R.}\ \bibnamefont
  {Nelson}}\ and\ \bibinfo {author} {\bibfnamefont {R.~A.}\ \bibnamefont
  {Pelcovits}},\ }\href@noop {} {\bibfield  {journal} {\bibinfo  {journal}
  {Phys. Rev. B}\ }\textbf {\bibinfo {volume} {16}},\ \bibinfo {pages} {2191}
  (\bibinfo {year} {1977})}\BibitemShut {NoStop}%
\bibitem [{\citenamefont {Chaikin}\ and\ \citenamefont
  {Lubensky}(2000)}]{chaikin2000principles}%
  \BibitemOpen
  \bibfield  {author} {\bibinfo {author} {\bibfnamefont {P.~M.}\ \bibnamefont
  {Chaikin}}\ and\ \bibinfo {author} {\bibfnamefont {T.~C.}\ \bibnamefont
  {Lubensky}},\ }\href@noop {} {\emph {\bibinfo {title} {Principles of
  condensed matter physics}}},\ Vol.~\bibinfo {volume} {1}\ (\bibinfo
  {publisher} {Cambridge Univ Press},\ \bibinfo {year} {2000})\BibitemShut
  {NoStop}%
\bibitem [{\citenamefont {Lansac}\ \emph {et~al.}(2003)\citenamefont {Lansac},
  \citenamefont {Maiti}, \citenamefont {Clark},\ and\ \citenamefont
  {Glaser}}]{Lansac2003}%
  \BibitemOpen
  \bibfield  {author} {\bibinfo {author} {\bibfnamefont {Y.}~\bibnamefont
  {Lansac}}, \bibinfo {author} {\bibfnamefont {P.~K.}\ \bibnamefont {Maiti}},
  \bibinfo {author} {\bibfnamefont {N.~A.}\ \bibnamefont {Clark}}, \ and\
  \bibinfo {author} {\bibfnamefont {M.~A.}\ \bibnamefont {Glaser}},\ }\href
  {\doibase 10.1103/PhysRevE.67.011703} {\bibfield  {journal} {\bibinfo
  {journal} {Phys. Rev. E}\ }\textbf {\bibinfo {volume} {67}},\ \bibinfo
  {pages} {011703} (\bibinfo {year} {2003})}\BibitemShut {NoStop}%
\bibitem [{\citenamefont {Toner}\ and\ \citenamefont
  {Nelson}(1981)}]{Toner1981}%
  \BibitemOpen
  \bibfield  {author} {\bibinfo {author} {\bibfnamefont {J.}~\bibnamefont
  {Toner}}\ and\ \bibinfo {author} {\bibfnamefont {D.}~\bibnamefont {Nelson}},\
  }\href {http://prb.aps.org/abstract/PRB/v23/i1/p316\_1} {\bibfield  {journal}
  {\bibinfo  {journal} {Phys. Rev. B}\ }\textbf {\bibinfo {volume} {23}},\
  \bibinfo {pages} {316} (\bibinfo {year} {1981})}\BibitemShut {NoStop}%
\bibitem [{\citenamefont {Etxebarria}\ and\ \citenamefont
  {Ros}(2008)}]{etxebarria2008bent}%
  \BibitemOpen
  \bibfield  {author} {\bibinfo {author} {\bibfnamefont {J.}~\bibnamefont
  {Etxebarria}}\ and\ \bibinfo {author} {\bibfnamefont {M.~B.}\ \bibnamefont
  {Ros}},\ }\href@noop {} {\bibfield  {journal} {\bibinfo  {journal} {J. Mater.
  Chem.}\ }\textbf {\bibinfo {volume} {18}},\ \bibinfo {pages} {2919} (\bibinfo
  {year} {2008})}\BibitemShut {NoStop}%
\bibitem [{\citenamefont {Pintre}\ \emph {et~al.}(2010)\citenamefont {Pintre},
  \citenamefont {Serrano}, \citenamefont {Ros}, \citenamefont
  {Mart{\'\i}nez-Perdiguero}, \citenamefont {Alonso}, \citenamefont {Ortega},
  \citenamefont {Folcia}, \citenamefont {Etxebarria}, \citenamefont
  {Alicante},\ and\ \citenamefont {Villacampa}}]{pintre2010bent}%
  \BibitemOpen
  \bibfield  {author} {\bibinfo {author} {\bibfnamefont {I.~C.}\ \bibnamefont
  {Pintre}}, \bibinfo {author} {\bibfnamefont {J.~L.}\ \bibnamefont {Serrano}},
  \bibinfo {author} {\bibfnamefont {M.~B.}\ \bibnamefont {Ros}}, \bibinfo
  {author} {\bibfnamefont {J.}~\bibnamefont {Mart{\'\i}nez-Perdiguero}},
  \bibinfo {author} {\bibfnamefont {I.}~\bibnamefont {Alonso}}, \bibinfo
  {author} {\bibfnamefont {J.}~\bibnamefont {Ortega}}, \bibinfo {author}
  {\bibfnamefont {C.~L.}\ \bibnamefont {Folcia}}, \bibinfo {author}
  {\bibfnamefont {J.}~\bibnamefont {Etxebarria}}, \bibinfo {author}
  {\bibfnamefont {R.}~\bibnamefont {Alicante}}, \ and\ \bibinfo {author}
  {\bibfnamefont {B.}~\bibnamefont {Villacampa}},\ }\href@noop {} {\bibfield
  {journal} {\bibinfo  {journal} {J. Mater. Chem.}\ }\textbf {\bibinfo {volume}
  {20}},\ \bibinfo {pages} {2965} (\bibinfo {year} {2010})}\BibitemShut
  {NoStop}%
\end{thebibliography}

%

\end{document}